# An Optimised Flow for Futures: From Theory to Practice

Nicolas Chappe[a], Ludovic Henrio[a], Amaury Maillé[a], Matthieu Moy[a], and Hadrien Renaud[a,b]

a    Univ Lyon, EnsL, UCBL, CNRS, Inria, LIP, F-69342, LYON Cedex 07, France
b    École Polytechnique/Institut Polytechnique de Paris, Palaiseau, France

**Abstract**    A future is an entity representing the result of an ongoing computation. A synchronisation with a "get" operation blocks the caller until the computation is over, to return the corresponding value. When a computation in charge of fulfilling a future delegates part of its processing to another task, mainstream languages return nested futures, and several "get" operations are needed to retrieve the computed value (we call such futures "control-flow futures"). Several approaches were proposed to tackle this issues: the "forward" construct, that allows the programmer to make delegation explicit and avoid nested futures, and "data-flow explicit futures" which natively collapse nested futures into plain futures.

This paper supports the claim that data-flow explicit futures form a powerful set of language primitives, on top of which other approaches can be built. We prove the equivalence, in the context of data-flow explicit futures, between the "forward" construct and classical "return" from functions. The proof relies on a branching bisimulation between a program using "forward" and its "return" counterpart. This result allows language designers to consider "forward" as an optimisation directive rather than as a language primitive.

Following the principles of the Godot system, we provide a library implementation of control-flow futures, based on data-flow explicit futures implemented in the compiler. This small library supports the claim that the implementation of classical futures based on data-flow ones is easier than the opposite. Our benchmarks show the viability of the approach from a performance point of view.

**ACM CCS 2012**

- **Computing methodologies** → *Distributed programming languages*; **Parallel programming languages**;
- **Theory of computation** → Program semantics;

## The Art, Science, and Engineering of Programming

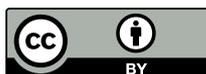

**An Optimised Flow for Futures: From Theory to Practice**

# 1 Introduction

A future [4] is an entity representing the result of an ongoing computation. It is used when launching a sub-task in parallel with the current task to later retrieve the result computed by the sub-task. A future is *resolved* when the associated value is computed. Futures have been used for more than 30 years now, and their adoption became wider and reached mainstream object-oriented languages during the last 20 years. Futures are used as a high-level parallelism paradigm in the standard libraries of Java, Scala, JavaScript, C++, or in the Akka and ProActive toolkits. A future is at the same time a container for some data and a way to synchronise processes: programs that use futures wait for the resolution of the future before executing some piece of code.

To classify futures, the classical approach is to distinguish implicit from explicit futures [7, 28], i.e whether there are dedicated operations to access a future's content (explicit) or not (implicit). We demonstrated [12, 17] that the way synchronisation is performed is a more distinctive feature. In particular futures should be classified depending on whether they are equipped with a data-flow or a control-flow synchronisation. We call the synchronisation "control-flow" when the execution of a given statement releases the synchronisation, and "data-flow" when it is the availability of some data that releases the synchronisation; such a distinction appears when nesting futures (see next paragraph). Control-flow futures are explicit by nature and generally typed using a parametric type of the form Fut[T]. Traditionally, data-flow futures have been implicit, but we showed they could be explicit as well. The typing of data-flow explicit futures requires specific typing rules that prevent the construction of nested future types. We call data-flow explicit futures *flows* [12] and denote their type Flow[T]. The type system for flows relies on a collapse rule that transforms a type of the form Flow[Flow[T]] into Flow[T].

**Delegation and Future Nesting** The distinction between control-flow and data-flow synchronisation makes a particular sense when the asynchronous task responsible for resolving a future delegates the resolution of this future to another asynchronous task (i.e. terminates its execution with a return *asynchronous-call*) and provides another future as the computed value. With *control-flow futures*, two synchronisations (e.g. get(get(nested_future))) are necessary to fetch the result: one waiting for the task that delegates and another for the task that resolves the future. With *data-flow* futures, a single synchronisation (e.g. get*(nested_future)) is necessary to wait for the resolution of both tasks. See listing 1.b on page 5 for an example of such a delegation. It is also in case of delegation that the difference between type systems is visible: the type of the result of an asynchronous call itself delegated asynchronously to another entity will be of the form Fut[Fut[int]]; with data-flow explicit futures such a type is *collapsed* into Flow[int].

Encore [9] is a language with control-flow futures but a focus on data-flow like synchronisations. Encore is an active object language extended with a forward construct allowing the programmer to encode delegation without creating nested future types [11]. The direct semantics of forward relies on future chaining but an optimising compilation phase avoids chaining and uses *promises*. Promises are variants of futures that are explicitly resolved through a `set` operation while a future is resolved when the associated task terminates.

**Contribution**  Compared to the previous works, this article further investigates the use of data-flow explicit futures and illustrates the benefits of this new kind of futures over existing ones. More precisely:

- We present both a formal definition and a concrete implementation of data-flow explicit futures in the Encore language.
- We illustrate the expressiveness and ease of programming brought by data-flow explicit futures both with illustrative examples and by showing that control-flow futures can easily be encoded from data-flow explicit futures; as such we provide the first complete implementation of an encoding of the Godot approach [12].
- The formal aspects of this article are based on `DeF`, a new core calculus dedicated to the study of data-flow explicit futures. `DeF` is not meant to be a new programming language, but rather a minimalistic formal language, expressive enough to be representative of existing mainstream languages. Because we target mainstream languages with imperative aspects, we define this calculus with a mutable state and some standard imperative constructs. It also features asynchronous calls that take advantage of data-flow explicit futures. The main improvement over Godot, a lambda-based calculus with data-flow explicit futures [12], is our support for mutable state.
- We formally and experimentally study one of the most promising optimisations enabled by data-flow futures: with data-flow explicit futures, the `return` and the `forward` primitives have the same semantics. We prove this equivalence and investigate the benefits we could obtain by exploiting it. Our calculus includes the formalisation of the `forward*` primitive with a novel typing rule that is safe even when the function that performs a `forward*` is invoked synchronously.

**Organisation of the Paper**  Section 2 presents existing future constructs and illustrates the benefits of data-flow explicit futures on an example. Section 3 presents `DeF`, a core calculus of data-flow explicit futures, its syntax, semantics, and type system. The extension of `DeF` with a `forward*` primitive is called `DeF+F`, it is also defined in this section. Section 4 proves that the `return` and the `forward*` primitives have the same behaviour by exhibiting a branching bisimulation between programs that use the `forward*` primitive and programs that only use `return`. Section 5 presents our `DeF` extension to Encore, the encoding of control-flow futures based on data-flow explicit futures, the implementation of the `forward*` primitive and concludes with a performance evaluation of the constructs presented in this paper.

## 2 Context and Related Work

### 2.1 A Brief History of Futures

"*Futures*" are a programming abstraction that has been introduced by [4]. It has then been used in programming languages like MultiLisp [16]. Futures provide an elegant form of parallelism by enabling the creation of asynchronous tasks that return a result. The synchronisation mechanism provided by futures is closely related with the flow of data in the programs. In MultiLisp, the future construct creates a thread and returns a future that can be manipulated by operations like assignment that do not need a real value, but the program would *automatically block* when performing an operation that requires the future value (e.g. an arithmetic operation). In MultiLisp, futures are implicitly accessed in the sense that there is no specific instruction for accessing a future but there is an explicit statement for creating them. Typed futures appeared with ABCL/f [27] to represent the result of asynchronous method invocations, i.e. invocations performed in parallel with the code that triggered them.

The first work on formalisation of futures was probably realised by Flanagan and Felleissen [13, 14] and was already focused on the difference between explicit and implicit future access. This work studies the futures of MultiLisp, that are explicitly created but implicitly accessed. The authors translate a program with futures into a lower-level program that explicitly retrieves futures, and then optimise the number of necessary future retrievals. In a similar vein, $\lambda(fut)$ [24] is a concurrent lambda calculus with futures with cells and handles. Futures in $\lambda(fut)$ are explicitly created, similarly to MultiLisp. Alice ML [23] can be considered as an implementation of $\lambda(fut)$.

While the first adoptions of futures were in functional languages, they are nowadays used in imperative and object-oriented programming. For example, futures now exist in many mainstream languages such as C++, Java, Javascript, Scala, ...In particular futures play a central role in actor [2] and active object languages [7], which are based on asynchronous communications between single-threaded entities. They use futures to represent replies to asynchronous messages. Except for ProActive [3], actor and active-object languages use explicit futures [9, 20, 21, 29]. In this context, De Boer et al. [8] designed a compositional proof theory for active objects, that puts a strong emphasis on the modelling of futures.

### 2.2 Motivational Examples

Listing 1 shows the same simple program in different languages and using different constructs: with control-flow explicit futures in Encore, with data-flow explicit futures in our extension to Encore, and with implicit futures in ProActive. The syntax of these languages is mostly simple object oriented programming except the following: in Encore, ! triggers an asynchronous method invocation that returns a control-flow explicit future; !! is the equivalent operation that creates data-flow futures; Fut[T] (resp. Flow[T]) is the type of a control-flow explicit future (resp. a data-flow explicit future) that will be resolved by a value of type T; get (resp. get*) waits for the resolution of a control-flow explicit future (resp. a data-flow explicit future) and returns its value.

Nicolas Chappe, Ludovic Henrio, Amaury Maillé, Matthieu Moy, and Hadrien Renaud

■ **Listing 1** A simple actor example: control-flow explicit futures, data-flow explicit futures and implicit futures (that are always data-flow).

a) Encore (control-flow explicit futures)

```
1  active class B
2    def bar(t: int): int
3      t * 2
4    end
5
6    def foo(x: int): Fut[int]
7      val t = x + 1
8      val beta = new B()
9      beta!bar(t)
10   end
11
12   -- we need this function as foo cannot take
          ↪ both fut and int
13   def foo_fut(x: Fut[int]): Fut[int]
14     this.foo(get(x))
15   end
16 end
17
18 active class Main
19   def main(): unit
20     val alpha = new B()
21     -- Asynchronous call using !, returns a
          ↪ future
22     val x: Fut[Fut[int]] = alpha!foo(1)
23     val y: int = get(get(x)) + 1
24     val z: Fut[Fut[int]] = alpha!foo_fut(get(x))
25     println(get(get(z))) -- 10
26   end
27 end
```

b) Encore with DeF (our extension)

```
1  active class B
2    def bar(t: int): int
3      t * 2
4    end
5
6    def foo(x: Flow[int]): Flow[int]
7      val t = get*(x) + 1
8      val beta = new B()
9      beta!!bar(t)
10   end
11 end
12
13 active class Main
14   def main(): unit
15     val alpha = new B()
16     -- Asynchronous call using !!, returns a flow
17     val x: Flow[int] = alpha!!foo(1) -- this lifts 1
          ↪ from int to Flow[int]
18     var y: int = get*(x) + 1
19     val z: Flow[int] = alpha!!foo(x)
20     println(get*(z)) -- 10
21   end
22 end
```

c) ProActive, implicit futures (Complete code for this example can be found on this gist)

```
1  public class B {
2      // ... define newB, main ...
3
4      public WrappedInt bar(WrappedInt t) {
5          return t.mult(2);
6      }
7
8      public WrappedInt foo(WrappedInt x) throws Exception {
9          WrappedInt t = x.add(1);
10         B beta = newB();
11         return beta.bar(t);
12     }
13
14     public static void realMain() throws Exception {
15         B alpha = newB();
16         WrappedInt x = alpha.foo(new WrappedInt(1)); // Method call on active object ⇒ asynchronous
                 ↪ call, returns an implicit future implementing the WrappedInt interface
17         WrappedInt y = x.add(1); // Forces synchronisation
18         WrappedInt z = alpha.foo(x);
19         System.out.println(z); // Forces synchronisation
20     }
21 }
```

In this article and in our implementation we use two different operators `get` and `get*` to distinguish operations on control-flow explicit futures and on data-flow explicit futures. Using polymorphism to offer a single primitive would be possible but we believe it is better to expose the programmer to two different operators that have a different semantics.

The example creates an actor `alpha` that is invoked asynchronously to perform the operation `foo`. The operation is partially delegated to an actor named `beta` as the result of foo is an asynchronous call to the method `bar` of `beta`. Notice the nested future type that exists with control-flow explicit futures (listing 1.a line 22) due to the delegated call and the subsequent double `get` operation line 23. With data-flow explicit futures, there are never two nested `Flow` types and never two consecutive invocations of `get*`. We call these futures *data-flow* because `get*` always returns data, never a future, but they are still explicit in the sense that the user needs to call `get*` akin to `get` in control-flow explicit futures. In ProActive, futures are implicit and the synchronisation occurs automatically on line 17 when an operation is triggered on the future. Implicit futures are always data-flow: because the type and the way to access data are transparent to the user, neither nested future types nor double-get operations would make sense for them. Finally, the example illustrates the invocation of a function with a future as parameter. With control-flow explicit futures the fact that a parameter is a future is explicit in a method's signature, but with data-flow explicit futures it is not necessary to declare two methods foo as an integer can be passed as parameter where a `Flow[int]` is expected. The synchronisation on line 7 of listing 1.b is resolved immediately if the parameter x is in fact an integer (that has been lifted to a `Flow`).

This example shows that the data-flow explicit futures are a good middle point between implicit futures and control-flow explicit futures, where the user keeps control over the synchronisation but delegates the chain handling to the language. The type system with data-flow explicit futures avoids the need for method duplication by automatically lifting native types to Futures. We now define more formally the different existing kinds of Futures, and position our work accordingly.

**2.3 The Limitations of Existing Future and Promise Constructs**

Two initial works highlighted the differences between explicit and implicit futures. First, it has been shown that adapting a static analysis [15] from explicit to implicit futures is difficult mostly because, with implicit futures, an unbounded number of synchronisation can be triggered by a given statement. Also the proof of correctness of the ProActive backend for ABS revealed the impossibility to simulate exhaustively control-flow synchronisation on implicit futures [19], this was due to the fact that synchronisation on implicit and explicit futures cannot observe the same thing.

Following this, a first active object version of data-flow explicit futures was proposed in a report [17]. Then the Godot system [12] was designed. It includes both data-flow and control-flow explicit futures and illustrates how to encode one into the other. The Godot system highlights three shortcomings of standard explicit futures and offers solutions:

Nicolas Chappe, Ludovic Henrio, Amaury Maillé, Matthieu Moy, and Hadrien Renaud

**The Future Type Proliferation Problem** leading to the nesting of future types in case of delegated calls. Line 22 of listing 1.a shows a nesting of future types. Data-flow explicit futures solve this problem.

**The Future Reference Proliferation Problem** referring to the possibly long chain of future references that has to be followed to reach the resolved future with either implicit or explicit futures in case of delegated calls. The use of forward can solve this problem.

**The Fulfilment Observation Problem** referring to the fact that the events observed with data-flow and with control-flow synchronisations are not the same. Indeed control-flow futures are better adapted to observe the flow of computation, e.g. in a scheduler, whereas data-flow futures are better adapted to express a computation and its data dependencies, e.g. in the resolution of a computational problem.

The current limitations of the Godot system are:

- The provided semantics is based on a stateless calculus while most of the languages that use futures are stateful. In particular the semantics does not allow a program to create a cycle of futures. While not desirable in practice, such cycles can happen as a consequence of the fact that the host language allows side effects. Our calculus is stateful and thus more representative of the existing implementations of futures.
- Incomplete implementation: the artefact associated with [12] encodes data-flow futures as a library built on top of the control-flow futures of Scala. It cannot handle parametric types that contain futures. Technically, the collapse rule is implemented at the time of creation of Flow to avoid creating a Flow[Flow[T]], but a Flow[Flow[T]] can be created by using a parametric type Foo[T] and instantiating it with T=Flow[T']. In this case the compiler has no way to collapse such types, and the type proliferation problem remains. To the best of our knowledge, no solution to this problem can work in a library-based approach without specific language support.
  We perform the reverse encoding (modify the language and compiler to implement data-flow natively, and encode control-flow on top of data-flow) and fully supports parametric types. The encoding of control-flow futures on top of data-flow ones was defined formally in [12], but was not implemented and therefore not studied experimentally.
- The typing rule of [12] for forward* allows calling synchronously a method that actually runs asynchronously, adding an implicit synchronisation. We believe this implicit synchronisation contradicts the principles of explicit futures, and as such, we provide a safer typing rule for forward*.

**About Promises** A promise is a future plus a handler that must be invoked to resolve the future. The handler is created at the same time as the future and any process that knows the handler can resolve the future. The advantage is that the resolution of the future is not tied to a given process. Promises do not suffer from the limitations of futures concerning delegation but they are more difficult to program. Indeed it is in general not possible to ensure that a promise is resolved exactly once [1]. Data-flow futures, explicit or implicit, keep the single resolution guarantee of futures while changing dynamically the thread responsible for fulfilling a future, which is the major advantage of promises compared to futures. Promises are used internally in Encore [11] to implement the forward primitive efficiently while avoiding the risks mentioned above.

**About Non-Blocking Future Access**  Another future access method is to register a continuation that will be executed asynchronously upon future resolution. It exists for example in Creol and AmbientTalk [10], where futures can only be accessed asynchronously, i.e. the programmer can only register some piece of code that will be executed when the future is resolved. Creol interrupts the execution of the current thread while waiting for the future. In AmbientTalk, *when-becomes-catch* behaves similarly but also a method invocation on a future generates an asynchronous invocation that will be scheduled when the future is resolved. In Akka, blocking future access is possible (using *Await.result* or *Await.ready*) but asynchronous future accesses should be preferred according to the documentation. Asynchronous reaction on future resolution is called *future chaining* in Encore [9], where it is provided by a ~~> operator (also called then). This operator registers code to be executed when the future is resolved; for example, with fut.then(lambda x ...) the resolution of fut triggers the execution of the lambda-expression on the right.

The advantage of non-blocking future accesses is that they prevent deadlocks. The counterpart is that the absence of synchronisation instruction makes it difficult to reason on the computations that are finished or not at a given program point. Many complete programming languages like Encore, ABS, and Akka feature both blocking and non-blocking future access.

Non-blocking access to data-flow futures makes perfect sense and, while our formalisation focuses on blocking synchronisation on futures, our results can be extended to non-blocking future access. On the practical side, we implemented non-blocking future access for data-flow futures in Encore (called chaining and denoted ~~>*).

## 2.4 Positioning and Discussion

The traditional classification of futures differentiates explicit from implicit futures [7, 28]. What is generally meant by the explicit vs implicit classification is the existence of 1) a specific type construct for futures (e.g., fut[int]), and 2) a specific future access primitive (e.g., get). In this sense our futures are explicit. However in all mainstream implementations of explicit futures, the type system for futures relies on classical parametric types that enforce a control-flow synchronisation on futures. Data-flow explicit futures are explicit but with a type system and synchronisation semantics that are different from usual explicit futures that use parametric types. Specifically, the type system of data-flow explicit futures *collapses* Flow[Flow[T]] into Flow[T] and it can *lift* T to Flow[T]. *Collapsing* enables data-flow synchronisation on futures while *lifting* allows a non-future type to be considered as a future.

One major example of a library with data-flow synchronisation on futures is the ProActive library [3]. GCM components [5], and a content addressable network [25] are two major projects that use the library. They both rely on delegation to route messages and on data-flow synchronisation for the access to results of invocations.

Data-flow synchronisation is more adapted to the programming of computations manipulating values. Control-flow synchronisation is better adapted to control explicitly the scheduling of tasks. Overall, control-flow explicit futures and data-flow explicit futures should probably coexist in a good programming library [12]. A short summary of the kinds of futures and their typical uses is given in table 1.

**Table 1** Summary of kinds of futures

| Suitable for / Benefits | | Explicit | Implicit |
|---|---|---|---|
| | | Easier debugging | Concise |
| Control-flow | task schedulers, … | C++, Java, Scala, Encore | *impossible* |
| Data-flow | computation using data | Godot, our contribution | MultiLisp, ProActive |

This article presents our modification to the Encore compiler to add support for data-flow explicit futures in addition to the existing explicit futures. We chose Encore because it is a full-fledged language and its compiler and type-checker are not too big. Moreover, Encore is an actor language where it is simple to create small examples that illustrate the delegation of asynchronous invocations and data-flow explicit futures.

This article also proposes an encoding of control-flow synchronisation based on data-flow explicit futures. This direction is the opposite one compared to the artefact of Godot [12], i.e. a Scala library that provides data-flow explicit futures based on existing control-flow explicit futures. Our encoding implementation is also more complete as it can fully deal with parametric types but its syntax is less nice: we only provide a set of functions and have no way to extend the syntax because our implementation is an Encore library. By implementing control-flow explicit futures as a library in a language that only provides data-flow explicit ones, the compiler has a single future primitive and the programmer can use both kinds of futures. We believe this approach is adapted to most languages that use explicit futures or promises.

Implementation matters are further discussed in section 5.

## 3 Data-Flow Explicit Futures: Principles and Semantics

This section presents two core languages called `DeF` (for data-flow explicit futures) and `DeF+F` that extends `DeF` with a `forward*` operator. `DeF` features data-flow explicit futures exclusively, functions that can be called synchronously or asynchronously, and a global state that enables imperative programming. We designed `DeF` as a minimalistic calculus but expressive enough so that our results on data-flow futures would still be relevant on a wide range of more complex calculi.

Our languages are equipped with a type system. Compared to our implementation of data-flow explicit futures in Encore we do not encode objects or actors and con-

▪ **Table 2** Static syntax of DeF.

$$
\begin{array}{rclr}
P & ::= & \overline{T\ x}\ \overline{M}\ \{\overline{T\ x}\ s\} & \text{program} \\
M & ::= & T\ \mathtt{m}(\overline{T\ x})\ \{\overline{T\ x}\ s\} & \text{function} \\
s & ::= & \mathtt{skip}\ |\ x = z\ |\ \mathtt{if}\ v\ \{s\}\ \mathtt{else}\ \{s\}\ |\ s\,;s\ |\ \mathtt{return}\ v & \text{statements} \\
z & ::= & e\ |\ \mathtt{m}(\overline{v})\ |\ \mathtt{!m}(\overline{v})\ |\ \mathtt{get}*\,v & \text{right-hand-side of assignments} \\
e & ::= & v\ |\ v \oplus v & \text{expressions} \\
v & ::= & x\ |\ \textit{integer-and-boolean-values} & \text{atoms} \\
B & ::= & \mathtt{Int}\ |\ \mathtt{Bool} & \text{basic type} \\
T & ::= & B\ |\ \mathtt{Flow}[B] & \text{Type}
\end{array}
$$

▪ **Table 3** Runtime Syntax of DeF.

$$
\begin{array}{rclr}
cn & ::= & a\,\rangle\,F & \text{configuration} \\
F & ::= & f\,(\overline{q})\ f\,(w) & \text{set of futures in configuration (unresolved / resolved)} \\
q & ::= & \{\ell | s\} & \text{stack frame} \\
w & ::= & f\ |\ b & \text{runtime values: future identifiers and basic values} \\
b & ::= & \textit{integer-and-boolean-values} & \text{values of basic types} \\
\ell, a & ::= & [\overline{x} \mapsto \overline{w}] & \text{local and global store} \\
s & ::= & \mathtt{skip}\ |\ x = z\ |\ \mathtt{if}\ v\ \{s\}\ \mathtt{else}\ \{s\}\ |\ s\,;s\ |\ \mathtt{return}\ v & \text{statements} \\
v & ::= & x\ |\ w & \text{variable or runtime value} \\
e & ::= & v\ |\ v \oplus v & \text{expressions with runtime values} \\
z & ::= & e\ |\ \mathtt{m}(\overline{v})\ |\ \mathtt{!m}(\overline{v})\ |\ \mathtt{get}*\,v & \text{right hand side of assignments}
\end{array}
$$

sequently data-races exist in DeF but not in Encore. We show in section 5.3 how to implement control-flow explicit futures on top of DeF.

### 3.1 Syntax of DeF

We use the following notations in our syntax. A bar over an expression, e.g. $\overline{q}$ denotes a list. All lists are ordered, except the set of futures in a configuration. $\varnothing$ is the empty list, and $q\#\overline{q}$ is the ordered list $\overline{q}$ with $q$ prepended to it. As for sets of futures, $FF'$ simply denotes the union of the sets $F$ and $F'$. $\oplus$ denotes any usual integer or boolean binary operator. Table 2 shows the static syntax of DeF. A program $P$ is made of a list of typed global variable declarations, a list of function definitions, and a main function ($s$ is the body of the main function). Each function $M$ has a return type, a name, a list of typed arguments, a list of typed local variables, and a statement that is the function body. Asynchronous function calls are supported via the $\mathtt{!m}(\overline{v})$ syntax. If $B$ is a basic type, $\mathtt{Flow}[B]$ denotes the type of a data-flow explicit future that is to be resolved by a value of type $B$.

Table 3 describes the runtime syntax of DeF. The configuration of a running DeF program contains a global store $a$, a set of resolved futures $f(w)$ and a set of unresolved futures $f(\overline{q})$ each associated with a running call stack $\overline{q}$. Each frame $q$ of a call stack contains a local store $\ell$ and a statement to be executed $s$. Each store is defined as a mapping from variable names to runtime values where runtime values $w$ are basic values $b$ and future identifiers $f$. Note that we also allow expressions and future values to contain future identifiers, this will be useful for evaluating get* statements.

To evaluate a program $P = \overline{T\ x}\ \overline{M}\ \{\overline{T'\ x'}\ s\}$ one must place it in an *initial configuration*. To do this we suppose that all variables have an initialisation value denoted[1] 0. The initial configuration for $P$ is: $a \rangle f_0(\{\ell|s\})$ where $f_0$ is any future identifier, the global store $a = [\overline{x \mapsto 0}]$ maps all global variables to an initialisation value, and $\ell = [\overline{x' \mapsto 0}]$ maps all local variables of the main body to an initialisation value.

The sequence $s\ ;\ s$ is associative and skip is neutral as the statement has no effect; thus we can rewrite any statement $s$ under the form $s'\ ;\ s''$ where $s'$ is not a sequence (and $s''$ might be skip if $s$ is a single statement). In the following we suppose that every statement is rewritten under this form (this simplifies the operational semantics).

Configurations are identified modulo reordering of futures (hence $Ff$ picks any $f$ in the set, not necessarily the last one) and future identifiers are unique; if $f(w) \in cn$ and $f(w') \in cn$, necessarily $w = w'$. Thus a configuration can be considered as a mapping from future identifiers to call stacks or values.

### 3.2 Semantics of DeF

Figure 1 details the small-step operational semantics of DeF that uses three notations:
- Similarly to other languages with binding of methods or functions [20], provided the program that is evaluated defines a function $T\ \mathrm{m}(\overline{T\ x})\ \{\overline{T\ y}\ s\}$, the bind operator instantiates a new stack frame with the local environment and the body of the function to be executed: $\quad \mathrm{bind}(m, \overline{w}) = \{[\overline{x \mapsto w}, \overline{y \mapsto 0}]\ |\ s\}$
- Given two stores $a$ and $\ell$, $(a + \ell)$ is the union of the two stores with values taken in $\ell$ in case of conflict: $(a + \ell)(x) = \ell(x)$ if $x \in \mathrm{dom}(\ell)$ and $(a + \ell)(x) = a(x)$ otherwise.[2]
- Given two stores $a$ and $\ell$, $(a + \ell)[x \mapsto w]$ is defined as $(a, \ell[x \mapsto w])$ if $x \in \mathrm{dom}(\ell)$, or $(a[x \mapsto w], \ell)$ otherwise.

Our semantics features asynchronous calls. INVK-ASYNC spawns an asynchronous task by adding an unresolved future to the configuration. From this point, the callee executes the spawned task in parallel with the caller. Once the spawned task is completed, the callee fulfils the future through the rule RETURN-ASYNC. The semantic rules for synchronous calls are more standard. INVK-SYNC pushes a new stack frame initialised in accordance with the function called and the arguments provided, and RETURN-SYNC pops the current stack frame and resumes the execution of the caller with the return value properly propagated.

The get* operator retrieves the value of a future $f$, defining the synchronisation points of a DeF program. Indeed, rules GET-FUTURE and GET-DATA are only enabled when getting a future of the form $f(w)$, that is, a fulfilled future. Consequently, performing a get* on an unresolved future blocks the process trying to access the future. Thus, this introduces a synchronisation when getting an unresolved future.

---

[1] Defining initial values for each existing type is not detailed here, note that 0 is a valid value for a Flow[int], corresponding to an already resolved future.

[2] We suppose that the program is type-checked and every variable is declared.

$$\overline{[\![w]\!]_\ell = w} \qquad \frac{x \in \mathrm{dom}(\ell)}{[\![x]\!]_\ell = \ell(x)} \qquad \frac{[\![v]\!]_\ell = k \quad [\![v']\!]_\ell = k'}{[\![v \oplus v']\!]_\ell = k \oplus k'} \qquad \textsc{Skip} \\ \overline{a \rangle F\ f(\{\ell \mid \texttt{skip}\,;s\}\#\overline{q})} \\ \rightarrow a \rangle F\ f(\{\ell \mid s\}\#\overline{q})$$

$$\textsc{Assign} \\ \frac{[\![e]\!]_{a+\ell} = w \quad (a+\ell)[x \mapsto w] = a' + \ell'}{a \rangle F\ f(\{\ell \mid x = e\,;s\}\#\overline{q})} \\ \rightarrow a' \rangle F\ f(\{\ell' \mid s\}\#\overline{q})$$

$$\textsc{Invk-Async} \\ \frac{[\![\overline{v}]\!]_{a+\ell} = \overline{w} \quad \mathrm{bind}(m,\overline{w}) = q' \quad f'\ \text{fresh}}{a \rangle F\ f(\{\ell \mid x =\,!m(\overline{v})\,;s\}\#\overline{q})} \\ \rightarrow a \rangle F\ f(\{\ell \mid x = f'\,;s\}\#\overline{q})\ f'(q')$$

$$\textsc{Invk-Sync} \\ \frac{[\![\overline{v}]\!]_{a+\ell} = \overline{w} \quad \mathrm{bind}(m,\overline{w}) = q'}{a \rangle F\ f(\{\ell \mid x = m(\overline{v})\,;s\}\#\overline{q})} \\ \rightarrow a \rangle F\ f(q'\#\{\ell \mid x = m(\overline{v})\,;s\}\#\overline{q})$$

$$\textsc{Return-Async} \\ \frac{[\![v]\!]_{a+\ell} = w}{a \rangle F\ f(\{\ell \mid \texttt{return}\ v\,;s\}) \rightarrow a \rangle F\ f(w)}$$

$$\textsc{Return-Sync} \\ \frac{[\![v]\!]_{a+\ell'} = w}{a \rangle F\ f(\{\ell' \mid \texttt{return}\ v\,;s\}\#\{\ell \mid x = m(\overline{v})\,;s'\}\#\overline{q})} \\ \rightarrow a \rangle F\ f(\{\ell \mid x = w\,;s'\}\#\overline{q})$$

$$\textsc{Get-Future} \\ \frac{[\![v]\!]_{a+\ell} = f'}{a \rangle F\ f(\{\ell \mid y = \texttt{get}\!\ast\ v\,;s\}\#\overline{q})\ f'(w')} \\ \rightarrow a \rangle F\ f(\{\ell \mid y = \texttt{get}\!\ast\ w'\,;s\}\#\overline{q})\ f'(w') \qquad \textsc{Get-Data} \\ \frac{[\![v]\!]_{a+\ell} = b}{a \rangle F\ f(\{\ell \mid y = \texttt{get}\!\ast\ v\,;s\}\#\overline{q})} \\ \rightarrow a \rangle F\ f(\{\ell \mid y = b\,;s\}\#\overline{q})$$

**Figure 1** Semantics of DeF (rules IF-TRUE and IF-FALSE for reducing if omitted). Recall that the set of futures $F$ in a configuration is not ordered.

Once the relevant future is fulfilled, repeated applications of GET-FUTURE will follow a sequence of futures and, unless there is a loop of futures or a deadlock, GET-DATA will finally provide the result of the get* operation.

Concretely, if there is a sequence of futures $f_0(f_1)\ldots f_{n-1}(f_n)\ f_n(w)$ in the configuration $cn$ (with $w$ not a future), a statement $y = \texttt{get}\ast\ f_0$ will become a $y = \texttt{get}\ast\ f_1$ statement thanks to the GET-FUTURE semantic rule, then $y = \texttt{get}\ast\ f_2$, and so on, until yielding a $y = \texttt{get}\ast\ w$ statement. At this point, the GET-DATA rule can be applied, reducing the statement to $y = w$. This resolution takes place at every get* statement: another $\texttt{get}\ast\ f_0$ will lead to the same series of GET-FUTURE applications. In this example DeF futures differ from control-flow explicit futures, as for the latter getting from $f_0$ to $w$ would have needed $n+1$ explicit get statements. When $n$ cannot be defined statically such a sequence is not expressible in the case of control-flow explicit futures. This makes sense as a sequence of $n+1$ futures is the result of $n+1$ levels of asynchronous delegations. Control-flow explicit futures follow the *control-flow* of the program while data-flow explicit futures follow its *data-flow*.

**Example** We show here a few examples of application of the semantics inspired by the Encore program in listing 1.b.

A function invocation corresponding to the call to *foo* can be expressed by the following reduction (applying the rule Invk-Async):

$$\varnothing \rangle f_0(\{\varnothing \mid x = !foo(1) \,;\, \dots \,;\, y = \texttt{get*}\ x\})$$
$$\to \varnothing \rangle f_0(\{\varnothing \mid x = f \,;\, \dots \,;\, y = \texttt{get*}\ x\})\ f(\{[x \mapsto 1] \mid \dots\})$$

Later on the resolution of future $f$ will be delegated to a *bar* function that returns 4. We reach the following configuration and apply the Return-Async rule:

$$\varnothing \rangle f_0(\{[x \mapsto f] \mid y = \texttt{get*}\ x\})\ f(\{[x \mapsto 1, y \mapsto f'] \mid \texttt{return}\ y\})\ f'(4)$$
$$\to \varnothing \rangle f_0(\{[x \mapsto f] \mid y = \texttt{get*}\ x\})\ f(f')\ f'(4)$$

Finally, the $y = \texttt{get*}\ x$ statement can be reduced, fetching the future value in three steps (2 Get-Future and one Get-Data)

$$\varnothing \rangle f_0(\{[x \mapsto f] \mid y = \texttt{get*}\ x\})\ f(f')\ f'(4) \to \varnothing \rangle f_0(\{[x \mapsto f] \mid y = \texttt{get*}\ f'\})\ f(f')\ f'(4)$$
$$\to \varnothing \rangle f_0(\{[x \mapsto f] \mid y = \texttt{get*}\ 4\})\ f(f')\ f'(4) \to \varnothing \rangle f_0(\{[x \mapsto f] \mid y = 4\})\ f(f')\ f'(4)$$

### 3.3 Syntax and Semantics of DeF+F

We now extend the language with a forward* statement. forward* takes a Flow as parameter and can be used instead of return to terminate the execution of a function. forward* f delegates the computation performed by the current task to the task that is computing f but only if forward* f is in the body of a function called asynchronously. If the function is called synchronously, forward* behaves like return. We present the consequences of this addition on syntax and semantics. The static syntax of DeF+F is the same as DeF plus a forward* statement:

$$s ::= \texttt{skip}\ \mid\ x = z\ \mid\ \texttt{if}\ v\ \{s\}\ \texttt{else}\ \{s\}\ \mid\ s\,;s\ \mid\ \texttt{return}\ v\ \mid\ \texttt{forward*}\ v$$

The runtime configurations of DeF+F have one more kind of future: chained futures.

$$F ::= \overline{f(\overline{q})}\ \overline{f(w)}\ \overline{f(\texttt{chain}\ f')}$$

Figure 2 defines the four semantic rules associated with forward*. Forward-Sync is similar to Return-Sync and allows using forward* with the same semantics as return in synchronous calls context, Forward-Data is similar to Return-Async but limited to the trivial case of a forward* of a non-future value.

The rule Forward-Async complements them by handling the forwarding of future values in an asynchronous context. Where Return-Async would have inserted an $f(f')$ into the context, Forward-Async inserts a $f(\texttt{chain}\ f')$ instead. Then, an application of Chain-Update will replace the chained future with a resolved future.

Concretely, if there is a sequence of futures $f_0(\texttt{chain}\ f_1)\dots f_{n-1}(\texttt{chain}\ f_n)\ f_n(w)$ in the configuration *cn*, Chain-Update will replace $f_{n-1}(\texttt{chain}\ f_n)$ with $f_{n-1}(w)$, then $f_{n-2}(\texttt{chain}\ f_{n-1})$ with $f_{n-2}(w)$, and so on. The *n*-th Chain-Update will update $f_0$ to $f_0(w)$. At this point, assuming $w$ is not a future, a get $f_0$ statement will only need a single Get-Future to reach a Get-Data transition.

FORWARD-ASYNC
$$\frac{[\![v]\!]_{a+\ell} = f'}{a \rangle F\, f(\{\ell \mid \text{forward*}\, v\,;\, s\}) \to a \rangle F\, f(\text{chain}\, f')}$$

FORWARD-SYNC
$$\frac{[\![v]\!]_{a+\ell} = w}{a \rangle F\, f(\{\ell \mid \text{forward*}\, v\,;\, s\}\#q\#\overline{q}) \to a \rangle F\, f(\{\ell \mid \text{return}\, w\,;\, s\}\#q\#\overline{q})}$$

FORWARD-DATA
$$\frac{[\![v]\!]_{a+\ell} = b}{a \rangle F\, f(\{\ell \mid \text{forward*}\, v\,;\, s\}) \to a \rangle F\, f(b)}$$

CHAIN-UPDATE
$$\frac{}{a \rangle F\, f(\text{chain}\, f')\, f'(w) \to a \rangle F\, f(w)\, f'(w)}$$

■ **Figure 2** Additional rules for the semantics of DeF+F.

The forward* construct proposes similar behaviour to return in DeF, but with operations occurring in a different order: in the case of forward*ed futures, future values are propagated as soon as possible, from the inner future to the outer one, and the resolution is performed only once per future, no matter how many delegated invocations there are. If return is used instead, several successive future retrieval operations occur until the inner future is reached.

**Example** We show here how a configuration evolves differently when forward* is used instead of return. For that purpose, as in section 3.2 we consider a configuration resulting from the execution of the Encore program in listing 1.b, slightly modified so that foo uses a forward* to return its result.

When the execution reaches line 9, if we are in a similar state as the example in the previous section we can apply the FORWARD-ASYNC rule:

$$\varnothing \rangle f_0(\{[x \mapsto f] \mid y = \text{get*}\, x\})\, f(\{[x \mapsto 1, y \mapsto f'] \mid \text{forward*}\, y\})\, f'(4)$$
$$\to \varnothing \rangle f_0(\{[x \mapsto f] \mid y = \text{get*}\, x\})\, f(\text{chain}\, f')\, f'(4)$$

Since $f'$ is a resolved future, CHAIN-UPDATE can be applied immediately:

$$\varnothing \rangle f_0(\{[x \mapsto f] \mid y = \text{get*}\, x\})\, f(\text{chain}\, f')\, f'(4)$$
$$\to \varnothing \rangle f_0(\{[x \mapsto f] \mid y = \text{get*}\, x\})\, f(4)\, f'(4)$$

Then, the $y = \text{get*}\, x$ statement can be reduced, fetching the future value with only one GET-FUTURE and one GET-DATA.

$$\varnothing \rangle f_0(\{[x \mapsto f] \mid y = \text{get*}\, x\})\, f(4)\, f'(4) \to \varnothing \rangle f_0(\{[x \mapsto f] \mid y = \text{get*}\, 4\})\, f(4)\, f'(4)$$
$$\to \varnothing \rangle f_0(\{[x \mapsto f] \mid y = 4\})\, f(4)\, f'(4)$$

Somehow, the CHAIN-UPDATE transition replaces one of the GET-FUTURE applications of the example in section 3.2. Section 4 will show that despite this difference, return and forward* have in fact equivalent semantics in DeF+F.

**3.4 The Type Systems of DeF and DeF+F**

Figure 3 gives a type system for DeF, and figure 4 introduces an additional typing rule for DeF+F. There are four kinds of type judgements: $\Gamma \vdash z : T$ types any expression $z$;

Type judgements for expressions:

$$\frac{}{\Gamma \vdash x : \Gamma(x)} \text{(T-Var)} \qquad \frac{\Gamma \vdash z : B}{\Gamma \vdash z : \text{Flow}[B]} \text{(T-Subtype)} \qquad \frac{\Gamma \vdash v : T \quad \Gamma \vdash v' : T' \quad \oplus : T \times T' \to T''}{\Gamma \vdash v \oplus v' : T''} \text{(T-Expression)}$$

$$\frac{\Gamma(\mathtt{m}) = \overline{T} \to T' \quad \Gamma \vdash \overline{v} : \overline{T}}{\Gamma \vdash !\mathtt{m}(\overline{v}) : \downarrow\!\text{Flow}[T']} \text{(T-Invk-Async)} \qquad \frac{\Gamma \vdash v : \text{Flow}[B]}{\Gamma \vdash \mathtt{get}* v : B} \text{(T-Get)} \qquad \frac{\Gamma(\mathtt{m}') = \overline{T} \to T' \quad \Gamma \vdash \overline{v} : \overline{T}}{\Gamma \vdash \mathtt{m}'(\overline{v}) : T'} \text{(T-Invk-Sync)}$$

Type judgements for statements:

$$\frac{\Gamma(x) = T \quad \Gamma \vdash e : T}{\Gamma \vdash_\mathtt{m} x = e} \text{(T-Assign)} \qquad \frac{\Gamma \vdash_\mathtt{m} s \quad \Gamma \vdash_\mathtt{m} s'}{\Gamma \vdash_\mathtt{m} s \,;\, s'} \text{(T-Seq)} \qquad \frac{\Gamma \vdash e : T' \quad \Gamma(\mathtt{m}) = \overline{T} \to T'}{\Gamma \vdash_\mathtt{m} \mathtt{return}\; e} \text{(T-Return)} \qquad \frac{}{\Gamma \vdash_\mathtt{m} \mathtt{skip}} \text{(T-Skip)}$$

Type judgements for programs and functions:

$$\frac{\Gamma' = \Gamma[\overline{x} \mapsto \overline{T}][\overline{x'} \mapsto \overline{T'}] \quad \Gamma' \vdash_\mathtt{m} s}{\Gamma \vdash T''\, \mathtt{m}\, (\overline{T\,x})\{\overline{T'\,x'}\; s\}} \text{(T-Method)} \qquad \frac{\Gamma = [\overline{x} \mapsto \overline{T}] \quad \Gamma[\overline{x'} \mapsto \overline{T'}] \vdash s \quad \forall M \in \overline{M}.\; \Gamma \vdash M}{\Gamma \vdash \overline{T\,x}\; \overline{M}\; \{\overline{T'\,x'}\; s\}} \text{(T-Program)}$$

**Figure 3** Type system; each operator $\oplus$ has a predefined signature, the rule for if is omitted.

$$\frac{\Gamma \vdash e : \text{Flow}[T'] \quad \Gamma(\mathtt{m}) = \overline{T} \to \text{Flow}[T']}{\Gamma \vdash_\mathtt{m} \mathtt{forward}* e} \text{(T-Forward)}$$

**Figure 4** Type system addition for DeF+F.

$\Gamma \vdash_\mathtt{m} s$ types a statement $s$ that belongs to the function m; $\Gamma \vdash T''\, \mathtt{m}\, (\overline{T\,x})\{\overline{T'\,x'}\; s\}$ checks that a function definition is well-typed; and $\Gamma \vdash \overline{T\,x}\; \overline{M}\; \{\overline{T'\,x'}\; s\}$ checks the correct typing of a program. Due to the fact that a flow already encodes an arbitrary number of asynchronous delegations of a computation, flows of flows are forbidden at the type system levels. The notation $\downarrow$ represents the collapsing of a flow type: $\downarrow\!\text{Flow}[\text{Flow}[T]]$ reduces to $\downarrow\!\text{Flow}[T]$. The interested reader should refer to [12] for a complete specification of collapse in a type system that supports parametric types. In our simplified context, the description can be summarised by the following rules:

$$\downarrow\!\text{Flow}[\text{Flow}[T]] = \downarrow\!\text{Flow}[T] \qquad \downarrow\!\text{Flow}[T] = \text{Flow}[\downarrow T] \text{ if } T \neq \text{Flow}[T'] \qquad \downarrow B = B$$

In figure 3, rule T-Subtype allows any basic type to be considered as a flow type. T-Invk-Async states that the type resulting of an asynchronous function invocation is flow containing the type returned by the function; the collapse operator prevents nested flow types. Symmetrically, T-Get states that the result of the synchronisation on a flow is necessarily a basic type. Indeed as flows on flows do not exist and basic types can be lifted to flow types, a get* operation can always be typed and always returns a basic type. Other type-checking rules are standard.

Concerning the rule T-FORWARD shown in figure 4, it types the forward* statement. It checks that the function's return type is Flow[$T'$] and that the returned type is compatible with this signature. This ensures that a synchronous call to a function that performs a forward* must consider the result of type Flow[$T$]. Our solution contrasts with what was adopted in Encore, where synchronously calling a function that contains a forward performs an implicit synchronisation on the future returned by the function; see Appendix B for more details. Because of the subtyping rule, $e$ could be of basic type $B$ but $T'$ cannot be of the form Flow[$T''$].

**Properties of the Type System**   The type system is not particularly original compared to the one of the Godot system [12] except the rule T-FORWARD that elegantly solves the typing of synchronous calls with forward, as explained above. It shares the classical properties: *progress* and *preservation* (reduction doesn't break typing). Both properties are expressed on DeF+F, and also valid on DeF which is a subset of DeF+F.

For stating and proving preservation one needs to extend the type system in order to type configurations. The extension raises no particular difficulty: each thread is type-checked separately, we check that the type of values in the store fits with the type of the declared variables, and that for each future $f$ of type Flow[T], $T$ is the type of the value stored in the future $f$ (or computed by the thread that computes $f$), and a get $f$ operation returns a value of type $T$. To type a runtime configuration, we need an extended typing environment of the form:

$$\Omega ::= \Gamma \quad \overline{f \mapsto \overline{\Gamma, m}} \quad \overline{f \mapsto T}$$

The first $\Gamma$ is the type of the global store, then for each not yet resolved future, the future identifier is mapped to a stack of $\Gamma, m$ where each $\Gamma$ is the typing environment that types the function body and $m$ is the function name that provides the returned type, and the type of the future for the last $m$ of the stack; finally we have a second mapping for resolved futures that maps each future identifier to the type of the future value. The initial configuration of a well-typed program is well-typed. Then we can state the preservation theorem:

**Theorem 1** (Preservation). *A well-typed configuration remains well-typed during reduction.*

$$\Omega \vdash a \mathbin{\rangle} F \;\land\; a \mathbin{\rangle} F \to a' \mathbin{\rangle} F' \implies \exists \Omega'.\, \Omega' \vdash a' \mathbin{\rangle} F'$$

$\Omega$ and $\Omega'$ are additionally constrained: the global $\Gamma$ is the same in $\Omega$ and $\Omega'$, and each future defined in $\Omega$ is also defined in $\Omega'$ with the same type.

Like Encore and most of the languages with futures, our language has imperative features. Thus, contrarily to the Godot system, it is possible to create cycles of processes where each process references the future that is to be resolved by the next process in the cycle. In imperative languages with futures, and therefore in DeF and DeF+F, it is possible to have deadlocks in such situations. We only ensure progress in the absence of such deadlocks. This restriction of the progress property is in fact an advantage of our model: it shows that DeF and DeF+F model faithfully the deadlocks that exist in mainstream languages with futures.

We introduce the *Unresolved* predicate that checks whether a future is unresolved. It is defined by: $Unresolved(f, F) = \not\exists w.\, f(w) \in F$. We use it to state a progress property:

**Theorem 2** (Progress). *In a runtime configuration, each element that is an unresolved future can progress except if it tries to perform a* get *on an unresolved future.*

$$\Omega \vdash a \rangle f(\overline{q})\ F \implies (\exists a'\ F'.\ a \rangle f(\overline{q})\ F \to a' \rangle F' \wedge f(\overline{q}) \notin F')$$
$$\vee\ (\exists v, \ell, y, s, \overline{q'}.\ \overline{q} = \{\ell \mid y = \mathtt{get}*\ v\ ;\ s\} \# \overline{q'} \wedge Unresolved(\llbracket v \rrbracket_{a+\ell}, F))$$

This theorem ensures that any chosen task can progress unless it is trying to access an unresolved future. Consequently, a configuration that cannot progress only consists of tasks blocked on a future access, which implies that there is a cycle of futures. More generally, any chosen task is able to terminate except if it depends (indirectly) on itself or on a task that contains a non-terminating computation (e.g. an infinite recursive call).

## 4 Forward: A Safe Optimisation in DeF

The previous section showed the different runtime semantics for return in DeF/DeF+F and forward* in DeF+F. This section proves that despite this difference, they can be used interchangeably in DeF+F, by showing formally that the use of forward* instead of return does not alter the semantics of a DeF+F program. This positions the choice between return and forward* in DeF+F as an optimisation matter rather than a semantics matter.

To prove this, we provide a simple translation from a DeF+F program to a DeF program that replaces forward* by return statements. We then show using a branching bisimulation that the translation does not modify the behaviour of the program.

**Note on Bisimulations.** Bisimulations allow to compare the semantics of asynchronous programs [22]. Given a relation $\mathcal{R}$, two transition systems are said to be strongly bisimilar if for two states related by $\mathcal{R}$, a series of transitions on either side can be associated to the same series of transitions on the other side, with each pair of intermediate states linked by $\mathcal{R}$. Most of the time a strong bisimulation is too strict to compare programs and one needs to consider that some of the transitions do not need to be simulated. A weak bisimulation relaxes the definition of strong bisimulation by allowing a subset of transitions to take place at any time even without an equivalent in the other transition system. These transitions are called non-observable, or $\tau$-transitions, and are useful to encode updates of internal state.

This relaxation has a cost: two weakly similar programs might have some different properties, e.g. concerning the presence of deadlocks. Branching bisimulation is a compromise between strong and weak bisimulation that guarantees more properties but allows the presence of non-observable transitions. With a branching bisimulation a program that performs a $\tau$-transition must remain in relation with the same states, guaranteeing that $\tau$-transitions have almost no effect on the program state. A divergence-branching simulation is another compromise that further requires an

$$
\begin{array}{ccc}
\mathscr{R}\text{-Id-Store} & \begin{array}{c}\mathscr{R}\text{-Id-Resolved}\\ cn_{\text{F}}\ \mathscr{R}\ cn_{\text{D}}\end{array} & \begin{array}{c}\mathscr{R}\text{-Forward-Async}\\ cn_{\text{F}}\ \mathscr{R}\ cn_{\text{D}}\end{array} \\
\overline{a\ \mathscr{R}\ a} & \overline{cn_{\text{F}}\ f(w)\ \mathscr{R}\ cn_{\text{D}}\ f(w)} & \overline{cn_{\text{F}}\ f(\texttt{chain}\ f')\ \mathscr{R}\ cn_{\text{D}}\ f(f')}
\end{array}
$$

$$
\begin{array}{cc}
\begin{array}{c}\mathscr{R}\text{-ForwardElim}\\ cn_{\text{F}}\ \mathscr{R}\ cn_{\text{D}}\end{array} & \begin{array}{c}\mathscr{R}\text{-Chain-Update}\\ cn_{\text{F}}\ f(\texttt{chain}f')\ f'(w)\ \mathscr{R}\ cn_{\text{D}}\end{array} \\
\overline{cn_{\text{F}}\ f(\{\ell\mid s\}\#\overline{q})\ \mathscr{R}\ cn_{\text{D}}\ f(\{\ell\mid [\![s]\!]_{fwdElim}\}\#\overline{[\![q]\!]_{fwdElim}})} & \overline{cn_{\text{F}}\ f(w)\ f'(w)\ \mathscr{R}\ cn_{\text{D}}}
\end{array}
$$

$$
\begin{array}{cc}
\mathscr{R}\text{-Get-Future-F} & \mathscr{R}\text{-Get-Future-D} \\
\dfrac{cn_{\text{F}}\ f(f')\ f''(\{\ell_{\text{F}}\mid y=\texttt{get}*\ f;s_{\text{F}}\}\#\overline{q})\ \mathscr{R}\ cn_{\text{D}}}{cn_{\text{F}}\ f(f')\ f''(\{\ell_{\text{F}}\mid y=\texttt{get}*\ f';s_{\text{F}}\}\#\overline{q})\ \mathscr{R}\ cn_{\text{D}}} & \dfrac{cn_{\text{F}}\ \mathscr{R}\ cn_{\text{D}}\ f(f')\ f''(\{\ell_{\text{D}}\mid y=\texttt{get}*\ f;s_{\text{D}}\}\#\overline{q})}{cn_{\text{F}}\ \mathscr{R}\ cn_{\text{D}}\ f(f')\ f''(\{\ell_{\text{D}}\mid y=\texttt{get}*\ f';s_{\text{D}}\}\#\overline{q})}
\end{array}
$$

**Figure 5** Relation between DeF+F configurations and DeF configurations.

infinite sequence of $\tau$-transitions on either side to correspond to an infinite sequence of $\tau$-transitions on the other side.

Below, we prove a branching bisimulation. Our semantics are too different to ensure divergence-branching simulation because of the different synchronisation strategies between return and forward*: getting the value of a future that is in a cycle will block on the DeF+F side, but enter an infinite sequence of $\tau$-transitions on the DeF side.

### 4.1 Translation from DeF+F to DeF and Program Equivalence

A DeF+F program can be translated into a plain DeF program using the semantics-preserving transformation $[\![\ ]\!]_{fwdElim}$ defined as follows:

$$[\![\texttt{forward}*\ v]\!]_{fwdElim} \triangleq \texttt{return}\ v$$

Terms other than forward* are unchanged. To prove that $[\![\ ]\!]_{fwdElim}$ actually preserves the semantics of a DeF+F program, we define in figure 5 a relation $\mathscr{R}$. It associates a DeF+F configuration $cn_{\text{F}}$ and a DeF configuration $cn_{\text{D}}$ that represents a similar state of execution. We prove that two configurations related by $\mathscr{R}$ are bisimilar. More precisely, the equivalence we prove is a branching bisimulation that does not observe the update of intermediate futures in a sequence of chained futures (i.e. considers GET-FUTURE and CHAIN-UPDATE as $\tau$-transitions) as the precise resolution status of a future is internal state that does not matter for the observable state of a program.

The trivial rules $\mathscr{R}$-ID-STORE and $\mathscr{R}$-ID-RESOLVED state that two identical configurations are similar. $\mathscr{R}$-FORWARDELIM deals with syntactic equality modulo forward* elimination, simply replacing forward* by return.

These rules are not sufficient, as CHAIN-UPDATEs can happen at any time on the DeF+F side, making the executions of a DeF+F program and its DeF counterpart slightly diverge. We still want these configurations to be related by $\mathscr{R}$, which will in fact be needed by the first item of theorem 3, as CHAIN-UPDATE is a $\tau$-transition.

$\mathscr{R}$-FORWARD-ASYNC and $\mathscr{R}$-CHAIN-UPDATE deal with the fact that some futures are chained and others are not. The rule $\mathscr{R}$-FORWARD-ASYNC states that chaining a future to another one, as the semantic rule FORWARD-ASYNC does, is semantically equivalent to fulfilling it with this same future, as Return-Async does. Rule $\mathcal{R}$-Chain-Update can be used to undo the future chaining operation.

As the resolution of futures is done in a different order, and possibly at a different time, when forward* is used instead of return, the rule $\mathcal{R}$-Chain-Update and both $\mathcal{R}$-Get-Future rules are needed to associate configurations in which some futures are at different stages of resolution. This different ordering only occurs upon resolution of the get* statement, and is handled by the two Get-Future rules.

This different ordering of future resolution also implies that there isn't a one-to-one mapping between Chain-Update transitions in a DeF+F execution and Get-Future transitions in the context of that same program after forward* elimination. All these facts are formalised by theorem 3.

### 4.2 Branching Bisimulation between DeF+F and DeF

**Theorem 3** (Correctness of the translation from DeF+F to DeF). *$\mathcal{R}$ is a branching bisimulation between the operational semantics of the DeF+F program P and the operational semantics of the DeF program $[\![P]\!]_{fwdElim}$.*

*Let R range over observable transitions. If $cn_F \mathcal{R} \, cn_D$ then:*

$$cn_F \xrightarrow{\tau}^* cn'_F \implies cn'_F \mathcal{R} \, cn_D \qquad\qquad cn_D \xrightarrow{\tau}^* cn'_D \implies cn_F \mathcal{R} \, cn'_D$$
$$cn_F \xrightarrow{R} cn'_F \implies \exists cn'_D . cn_D \xrightarrow{\tau}^* \xrightarrow{R} cn'_D \wedge cn'_F \mathcal{R} \, cn'_D$$
$$cn_D \xrightarrow{R} cn'_D \implies \exists cn'_F . cn_F \xrightarrow{\tau}^* \xrightarrow{R} cn'_F \wedge cn'_F \mathcal{R} \, cn'_D$$

*The transitions Get-Future and Chain-Update are non-observable, both of them are labelled $\tau$. The observable transitions Forward-Async and Forward-Data are labelled Return-Async, and Forward-Sync is labelled Return-Sync. All the other transitions are labelled with their original rule name.*

As the use of forward* does not change when futures are created but only how they are resolved, if $cn_F \mathcal{R} \, cn_D$ then the identifiers of the futures of $cn_F$ are exactly those of $cn_D$, e.g. we have $f \in cn_F \iff f \in cn_D$. However, the value of $f$ in $cn_F$ and the value of $f$ in $cn_D$ may differ. In the following, we will denote that some futures $f \in cn_F$ and $f' \in cn_D$ actually share the same identifier by $f = f'$ or $f \equiv f'$.

First, we introduce the notion of sequence of futures, and define a few lemmas on properties implied by $cn_F \mathcal{R} \, cn_D$. This will help to prove that $\mathcal{R}$ is a branching bisimulation.

**Definition 1** (Sequence of futures). *In a DeF or DeF+F configuration cn, $(f_0 \ldots f_n)$ such that $\forall i, f_i(f_{i+1}) \in cn \vee f_i(\text{chain } f_{i+1}) \in cn$ is called a sequence of futures.*

Given a future resolved or chained to another one in a DeF+F configuration, lemmas 1 and 2 state the possible forms of the corresponding future in an associated DeF configuration. Conversely, given a future resolved to another one in a DeF configuration, lemma 3 states the possible forms of the corresponding future in the associated DeF+F configuration. Lemma 5 generalises lemmas 1 and 2: given a sequence of futures in a DeF+F configuration, it states the possible forms of the corresponding sequence in DeF, while lemma 6 generalises lemma 3 in a similar manner. As for lemma 4, it formalizes that the local store and statements of a task are not altered by $[\![\cdot]\!]_{fwdElim}$, apart from get* statements that can express a different stage of future resolution.

**Lemma 1** (Matching a chained future). *If $cn_F \mathcal{R} cn_D$ and $f(\text{chain } f') \in cn_F$, then $f(f') \in cn_D$.*

The case of a future resolved on the DeF+F side is more complicated, as such a future can not only come from a return statement, but also from a CHAIN-UPDATE. The following lemma illustrates that the chain construct can flatten chains of futures.

**Lemma 2** (Resolved future in DeF+F). *If $cn_F \mathcal{R} cn_D$ and $f(w) \in cn_F$, then there exists $f_0(f_1) \ldots f_{n-1}(f_n) \in cn_D$ such that $f_0 \equiv f$, $f_n(w) \in cn_D$, and $\forall i, f_i(w) \in cn_F$.*

The previous two lemmas dealt with futures resolved on the DeF+F side, the next one deals with futures resolved on the DeF side.

**Lemma 3** (Resolved future in DeF). *If $cn_F \mathcal{R} cn_D$ and $f(w) \in cn_D$, then:*
- *Either $w$ is a future and $f(\text{chain } w) \in cn_F$.*
- *Or there exists $w'$ such that $f(w') \in cn_F$ and there exists $f_0(f_1) \ldots f_{n-1}(f_n) \in cn_D$ such that $f \equiv f_0$, $f_n(w') \in cn_D$, and $\forall i, f_i(w') \in cn_F$.*

Given a resolved or chained future in a DeF or DeF+F configuration, the previous lemmas gave the form of the corresponding future on the other side. The next lemma deals with the last case: futures not yet resolved, that still have a task attached to them. As the $[\![\cdot]\!]_{fwdElim}$ transformation is fairly simple, the local store and most statements will be identical on both sides, but get* statements may differ. Indeed, following a sequence of futures by the rule GET-FUTURE being non-observable, $\mathcal{R}$ has to relate get* statements at different stages of update. In this case, walking back the chain of GET-FUTURE leads to the same initial future.

**Lemma 4** (Matching tasks). *If $cn_F \mathcal{R} cn_D$, then $\exists s. f(\{l \mid s\}\#\overline{q}) \in cn_F$ if and only if $\exists s'. f(\{l \mid s'\}\#\overline{[\![q]\!]_{fwdElim}}) \in cn_D$. In this case:*
- *Either $s' = [\![s]\!]_{fwdElim}$.*
- *Or $s$ is of the form $y = \text{get}* w; s_1$, and $s'$ of the form $y = \text{get}* w'; [\![s_1]\!]_{fwdElim}$, with*

$$\exists w_0 \ldots w_n \in cn_F. \exists w'_0 \ldots w'_m \in cn_D \begin{cases} \forall i < n \; w_i(w_{i+1}) \in cn_F \\ \forall i < m \; w'_i(w'_{i+1}) \in cn_D \\ w_0 \equiv w'_0 \\ w_n = w \\ w'_m = w' \end{cases}$$

The two last lemmas are generalisations of the three first ones. They answer the question: given a sequence of futures, on the DeF side or on the DeF+F side, what can we say about the other side? Lemma 5 is about the DeF to DeF+F case, while lemma 6 handles the other direction.

▶ **Lemma 5** (Sequence of futures: DeF to DeF+F)**.**
*If $cn_F \mathcal{R} \, cn_D$ and $f_0(f_1)\ldots f_{n-1}(f_n)\, f_n(w) \in cn_D$ with $\nexists f \in cn_D.\, w = f$, then there exists $k_0 \leqslant \cdots \leqslant k_l$ such that for all $i < l$ either $f_{k_i}(f_{k_{i+1}}) \in cn_F$ or $f_{k_i}(\mathtt{chain}\ f_{k_{i+1}}) \in cn_F$ with $k_0 = 0$ and $k_l = n$.*

▶ **Lemma 6** (Sequence of futures: DeF+F to DeF)**.**
*If $cn_F \mathcal{R}\, cn_D$ and $f_0(f_1)\ldots f_{n-1}(f_n)\, f_n(w) \in cn_F$,
then there exists $f'_0(f'_1)\ldots f'_{l-1}(f'_l) \in cn_D$ such that $f'_0 = f_0$ and $f'_l(w) \in cn_D$.*

The proof of theorem 3 is done by induction and a classical case analysis on the reduction rule applied, proving that the bisimulation relation is maintained. Details of the proofs are provided in appendix A. The most interesting cases are the Get-Data rules, where a future is about to be resolved on one side, but there may still be multiple $\tau$-transitions needed to get to the resolution on the other side.

In this section we have shown that, with data-flow explicit futures, the behaviour of the forward* primitive is the same as the behaviour of a standard return. This highlights the fact that forward provides a form of data-flow synchronisation for control-flow explicit futures. More interestingly this shows that if necessary any return statement that returns a data-flow future could be compiled similarly to a forward* statement, which should in general improve performances. In the next section, we will first present our implementation, in Encore of DeF and how DeF can be used to encode classical futures. We will also evaluate the performance of different future constructs and different ways to use them. Our objective is both to evaluate the effectiveness of data-flow future synchronisation and the opportunities for automatic optimisation of future flows.

## 5 DeF in Practice

This section has four purposes:
1. Describe an implementation of data-flow explicit futures in the Encore language. This is considered as a basis for the implementation of DeF in another language.
2. Explain the choice we made in the implementation of forward* in our extension to the Encore language.[3]
3. Describe our implementation of explicit futures based on data-flow explicit futures, following the approach proposed in [12]. This is crucial in order to provide to the programmer the choice between control-flow and data-flow synchronisation (cf. fulfilment observation problem above).
4. Evaluate the performance of different implementation and synchronisation strategies for futures on programs that express various communication patterns.

---
[3] https://gitlab.inria.fr/lhenrio/encorewithdatafuts, visited on 2021-05-23.

An Optimised Flow for Futures: From Theory to Practice

**5.1 Implementation of** Flow

To implement data-flow synchronisation for explicit futures in a language, three steps are necessary: 1) adapt the type system, 2) deal with the creation of data-flow explicit futures at runtime, and 3) provide manipulation primitives for data-flow futures.

**5.1.1 Typing**
There are two aspects related to the typing of Flow[T] that require special care: the subtyping rule (T-Subtype) and the collapsing rules (operator ↓ in T-Invk-Async). The subtyping occurs when the programmer tries to substitute a value of type T (T ≠ Flow) where a value of type Flow[T] is expected: such a substitution is correct. As for the collapsing rules, when the typechecker encounters the type Flow[T], it must collapse it as per the collapsing rules established in [12], and summarised in 3.4. This ensures that nested flow types do not appear and that lifting of non flow values into flows is possible.

**5.1.2 Creation of** Flow
In DeF flows are created when a value of type T is substituted to a value of type Flow[T] (*lifting*) and when a function is called in an asynchronous way using the !! operator in DeF. In our extension to Encore, additionally to lifting and the !! operator, we provide an async* operator that allows asynchronously executing arbitrary code. Asynchronous invocation of methods and asynchronous execution of arbitrary code do not need special care compared to control-flow explicit futures: one should create an instance of Flow and set it to be resolved with the result of the asynchronous execution. Lifting requires to create an instance of Flow and immediately resolve it with the lifted value (no process creation nor context-switching is needed). Lifting the value into a Flow allows programmers to write functions that may return results of asynchronous invocations as well as values lifted to an instance of Flow, unified under a single type.

**5.1.3** Flow **Synchronisation with** get*
In figure 2, rules Get-Future and Get-Data are used to recursively traverse a chain of flows until a value is reached. Get-Data is only applicable on non-flow values. An implementation typically calls get* recursively (Get-Future) until a non-flow value is found and returns this value (Get-Data), but deciding whether to perform a recursive call requires that the information of whether a value is a flow be available at runtime. In the absence of parametric types, the information is even available statically as part of typing, making the implementation straightforward. In the presence of parametric types, in a compilation scheme using type erasure, the compiler cannot know statically whether a parametric type P[T] is instantiated with T being a flow or a non-flow type. The decision to make a recursive call or not then requires introspection facilities built into the runtime. In languages where no such introspection features are available, implementing flows requires compiler designers to add runtime information to flows to determine whether an instance is resolved with another flow or with a value. In Encore, the Pony runtime [26] adds introspection facilities that we use to check whether a Flow is resolved with a flow or with a non-flow value.

### 5.2 Implementation of forward* in Encore

The forward primitive is easy to type and understand when it is in the body of a function that is run asynchronously. However, defining a good semantics and type system in the context of Encore where methods can be called either synchronously or asynchronously needs more attention. For example • symbol was used in [11] "to prevent the use of forward in contexts where the expected task type is not clear". In our case, similar difficulties arise with the conjunction of synchronous invocation and forward* statement.

The rule in figure 4 provides an elegant and safe solution to this problem. However, the existing implementation of forward in the Encore language is more permissive. Our theoretical results of section 4 are based on the rule of figure 4 because it is the safest solution but our implementation is a bit more permissive concerning the typing of functions that perform a forward*: the return type of such functions does not need to be Flow[T]. This changes nothing when the function is called asynchronously. Instead, this choice entails an additional synchronisation (get*) performed before the forward statement when a function that performs a forward* is called synchronously. This changes a bit the semantics and thus our result on equivalence between return and forward is only valid *in the Encore implementation* if the method is called asynchronously. However as asynchronous invocations are identified syntactically, it would be easy to use forward* only in the asynchronously called functions. These aspects are further discussed in appendix B.

Because the more permissive semantics introduces an implicit synchronisation on an explicit future, in a freshly designed language we would advise to use data-flow futures with the typing of forward* taken from figure 4. In our implementation we chose a more permissive and less safe version to be consistent with the Encore ecosystem.

### 5.3 Encoding Fut[] from Flow[]

In this section, we build control-flow explicit futures on top of DeF, to assert its backward compatibility with existing systems. We show that a language implementing only Flow[] can build Fut[] as a library. In Encore, a library has no possibility to extend the syntax or introspect the rest of the code, we are thus limited to simple encodings. In practice, we provide an implementation of Fut[] that relies on our Flow[] construct of Encore and on our implementation of get∗, ⇝∗ (the natural extension to data-flow explicit futures of the ⇝ operator for future chaining), and async*.

**Definition** Godot [12] suggests a construction for such control-flow explicit futures:

$$\text{Fut}[\tau] ::= \Box \text{Flow}[\tau]$$

$\Box$ is called the "box" operator. It encapsulates its argument in a structure of a different type, whose only available operation is unbox, where $\text{unbox}(\Box\ x) = x$. Intuitively, the $\Box$ operator stops type collapsing: $\downarrow \text{Flow}[\text{Flow}[T]]$ reduces to $\downarrow \text{Flow}[T]$, but $\downarrow \text{Flow}[\Box \text{Flow}[T]]$ is not collapsed (it reduces to $\text{Flow}[\Box \text{Flow}[T]]$).

The corresponding operations follow:
$$\text{get}\ e ::= \text{get} * (\text{unbox}\ e) \qquad \text{then}(e, f) ::= \Box \text{then} * (\text{unbox}\ e, f)$$

**Implementation** We define the class Future[t] and the get_[4] function as shown in listing 2.

■ **Listing 2** Future[t] and get_ based on Flow and get*

```
1  read class Future[t]
2      val content: Flow[t]
3      def init(x: Flow[t]): unit
4          this.content = x
5      end
6  end
7
8  fun get_(y: Future[t]): t
9      return get*(y.content)
10 end
```

A restriction of our approach is that, in an Encore library, we cannot overload existing operators or functions. As such, we provide functions with the right behaviour, but not the desired name: get_, call_, await_, or async_. For the call operator (!) the problem is slightly more complex as functions are not sufficient and the syntax should be extended. In the current state, programmers should use call_(this!!foo(…)) when they mean this!foo(…). A similar adaptation is needed for forward. This restriction is only a matter of syntactic sugar.

**Summary** With this implementation of control-flow explicit futures on top of data-flow explicit futures, we showed that control-flow explicit futures can be provided as an extension of DeF, and as a simple library. It is interesting to note that Godot provides an implementation in Scala of data-flow synchronisation on top of control-flow explicit futures (the opposite of what we did here), but that implementation does not fully support parametric types and its extension to parametric types would be challenging. Consequently, we believe that implementing DeF directly in the compiler is more flexible, and thus we advise, in the design of future programming languages with explicit futures, to first implement a data-flow synchronisation, and then extend it with control-flow explicit futures.

## 5.4 Benchmarks

In this section, we provide a performance analysis of different implementations of futures[5]: the builtin control-flow explicit futures Fut of Encore, the data-flow explicit futures DeF that we introduced earlier, denoted as Flow, and the control-flow explicit futures that were built in the previous subsection, denoted as Fut on Flow. We analyse several programs, using chains of futures of different length, or different memory management patterns. Our focus is to analyse the performance of futures, not the efficiency of the parallelisation in Encore or the number of actors that can be created in the Encore system. This is why we do not compare ourself with other frameworks and our programs are mostly sequential or with little parallelism. Reducing parallelism allow us to avoid the complex interactions that exist between additional objects, concurrent garbage collection, memory contention: we only analyse the direct impact of different forms of futures.

---

[4] Underscore appended to avoid name collision with the get reserved keyword.
[5] All the code for those benchmarks is available at: https://gitlab.inria.fr/datafut/fut-on-flow, visisted on 2021-05-23.

All the benchmark results provided here are done on the same Dell XPS 13 9370, with a 8-core Intel Core i7-8550U and 16GiB of memory, running Ubuntu 20.04 and clang v10.0. Encore at version `ea5736869d2ac34cfbeee2be4a2988c819a215af` is run using the release configuration and the -O3 flag.

### 5.4.1 Test Programs

We evaluate performance on four examples, the first using non-nested futures, the next two stressing the implementation with long chains of futures, and the last one being closer to real-world workloads.

**Word Counter example.** The Encore compiler is shipped with tests, one is a word counter, from which we adapted this example. The WordCounter test dispatches asynchronous hash table insertions to 32 actors. We re-implemented the standard library module `Big.HashTable.Supervisorh` with `Flow`, with `Fut` on flow and with sequential execution (referred to as fun). There is no need here to use any forward, because there is no chain of futures. The original example used an optimisation that removed the use of futures: send one-way messages when the caller does not need the result. As this does not use futures, we disabled it in the `Fut` example, but still displayed its performance as `OneWay`.

**Recursive calls on a single actor: Ackermann.** We implemented the Ackermann function with recursive calls on a single actor. This program performs successive asynchronous invocations on the same entity (no parallelism, long delegation chains). In this benchmark, the chains of futures are of various lengths, because each call makes two recursive call, one which is forwarded and one for which it waits. Because we only have one actor, this wait cannot be a synchronisation, otherwise it would block the execution. We use here a cooperative yield (await) on the future before calling get. This problem does not arise with forward.

**Recursive calls on many actors without actor creation: recursive list summation.** We compute the sum of the numbers from 1 to $n$ by creating a linked list of actors with those numbers and then traversing the list with successive asynchronous tail recursive invocations. The measured time does not take into account list creation. There is here one long chain of futures of length $n$. When forward is used, as one future is delegated along the chain of futures, this sum is done in constant space.

**K-Means** This benchmark runs a fixed number of iterations of the k-means clustering algorithm. Every compute intensive step is delegated to a large number of actors, then synchronised. This does not use forward since there are no chained futures. However, the benchmark is interesting because it allows us to compare the performance of Encore's native control-flow explicit futures and our re-implementation of it on top of data-flow explicit ones, on a realistic workload.

### 5.4.2 Results and Discussion

The performance of each of the four benchmarks are shown in tables 4, 5, 6, and 7.

**An Optimised Flow for Futures: From Theory to Practice**

■ **Table 4** Running time of WordCounter (100 runs). The OneWay line corresponds to the version futures are optimised out by the compiler.

| Future used | Average running time (ms) | Std deviation | |
|---|---|---|---|
| OneWay | 103.4 ms | 4.6 ms | 4.4% |
| Fut | 185.5 ms | 6.5 ms | 3.5% |
| Flow | 184.6 ms | 6.7 ms | 3.6% |
| Fut on Flow | 190.4 ms | 7.6 ms | 4.0% |
| fun | 119.3 ms | 2.4 ms | 2.0% |

■ **Table 5** Running time of the Ackermann benchmark (arguments: 3, 4 – 100 iterations)

| Future used | Average running time | Std deviation | |
|---|---|---|---|
| Fut with forward | 27.86 ms | 1.2 ms | 4.3% |
| Fut on Flow with forward | 28.10 ms | 1.0 ms | 3.5% |
| Flow with forward* | 31.49 ms | 0.7 ms | 2.2% |
| Flow | 39.13 ms | 1.0 ms | 2.4% |
| Fut | 44.63 ms | 0.7 ms | 1.7% |
| Fut on Flow | 46.71 ms | 0.8 ms | 1.6% |

■ **Table 6** Average running time of the recursive list summation (10000 actors, 100 iterations)

| Future used | Average running time | Std deviation | |
|---|---|---|---|
| Fut with forward | 16.03 ms | 1.0 ms | 6.4% |
| Fut on Flow with forward | 16.30 ms | 1.0 ms | 6.3% |
| Flow with forward* | 16.14 ms | 0.9 ms | 5.8% |
| Flow | 26.02 ms | 1.3 ms | 5.1% |
| Fut | 46.87 ms | 3.9 ms | 8.4% |
| Fut on Flow | 47.64 ms | 3.8 ms | 8.0% |

■ **Table 7** Running time of 10 k-means steps (100 clusters, 10000 observations, 100 iterations)

| Future used | Average running time | Std deviation | |
|---|---|---|---|
| Flow | 468.8 ms | 114.7 ms | 24.5% |
| Fut on Flow | 494.9 ms | 125.8 ms | 25.4% |
| Fut | 500.7 ms | 66.8 ms | 13.3% |

The ackermann example has lower standard deviation than the other benchmarks, this might be caused by smaller future chains than the list summation example, or the absence of parallelism, unlike the WordCounter example.

The OneWay version of the WordCounter example is unsurprisingly far faster than the others. It allows evaluating the overhead of the future synchronisation over a plain one-way message. The sequential version is the slowest, showing that other versions do exploit parallelism even with a fine granularity of synchronisation.

The Ackermann and recursive list summation examples show the performance benefit of using forward or forward* over the plain Fut or Flow equivalent when there are delegated computations.

On all examples, Fut and Fut on Flow perform similarly, showing that a library-based implementation of Fut based on data-flow explicit futures reaches performance similar to a dedicated implementation within the compiler. We can notice a small overhead that can be explained by the additional "box" objects introduced by Fut on Flow.

Both Ackerman and the recursive list summation show the performance benefit of using Flow over Fut, since less synchronisation is needed (for Ackermann, using Flow allowed us to directly return a Flow without the need for a costly await). This difference disappears when using forward on both versions since no synchronisation is needed whatever the type of future.

The k-means benchmark models real-world workloads and has a larger standard deviation than the others. However the observed performances still show that adding data-flow explicit futures does not affect significantly the performance of futures.

In the Fut version of the benchmarks, forward explicitly changes the semantics to relax synchronisation. As shown in section 4, in the Flow version, it becomes an optimisation that does not change semantics. A clever compiler could have compiled the Flow version to Flow with forward* with the same semantics, and these benchmarks show that this is an interesting optimisation. An implementation with DeF and implicit optimisation of `return` into `forward*` (for long chains) would ally both the performance improvements of forward and the improved type handling of DeF.

## 6 Conclusion

We presented both theoretical and practical results showing that data-flow explicit futures form a relevant programming abstraction for parallel programming. Other constructs such as classical control-flow explicit futures can be implemented on top of them, and they allow optimisations that used to require ad-hoc keywords.

Our forward* construct has a safe static semantics that constrains functions performing a forward* to return a Flow type. As a consequence, even if the function is called synchronously, the semantics stays the same as return. On the contrary, the original Encore language has a specific semantics for synchronously called functions: they are forbidden in the formalisation but in practice, they are implemented with an additional synchronisation that may lead to deadlocks.

One of the advantages of data-flow synchronisation is the possibility to optimise the order of transmission of futures to the processes that use it, this had been partially studied in a distributed setting in [18] but the current article opens new opportunities, allowing the compiler and the runtime to explore compromises between pulling the results with recursive get*, pushing it based on forward*, or using the optimised compilation into promises provided by Encore [11].

**About Failures**  Dealing with failures in concurrent and distributed systems is difficult. While futures can be safely integrated inside recovery protocols [6], it is more difficult to deal asynchronously with failures or timeouts inside the application code because the point of usage of a future can be far from the point of invocation of the computation and thus traditional exception handling mechanisms are often not adapted. Like for asynchronous reaction to fulfilment (section 2.3), mechanisms similar to what exists for control-flow futures can be adapted for handling exceptions or failures with futures. In the current state of our study it does not seem that data-flow synchronisation helps with the difficulty of handling failures in an asynchronous manner. However, the data-flow synchronisation offers a different point of reaction to failure, closer to the use of the computation result, which can help the programmer handle the failure from a computational point of view.

## A Proofs

This section gathers the lemmas used to prove the bisimulation theorem, their proof, and the proof of the theorem itself.

### A.1 Proofs of Lemmas

**Lemma 1** (Matching a chained future). *If $cn_F \mathcal{R} cn_D$ and $f(\text{chain } f') \in cn_F$, then $f(f') \in cn_D$.*

*Proof.* This property can be proved by induction on $\mathcal{R}$: the only relevant rules are $\mathcal{R}$-ID-RESOLVED and $\mathcal{R}$-FORWARD-ASYNC, both trivial. Note that by construction there is no chain on the right hand side of $\mathcal{R}$. □

**Lemma 2** (Resolved future in DeF+F). *If $cn_F \mathcal{R} cn_D$ and $f(w) \in cn_F$, then there exists $f_0(f_1)\ldots f_{n-1}(f_n) \in cn_D$ such that $f_0 \equiv f$, $f_n(w) \in cn_D$, and $\forall i, f_i(w) \in cn_F$.*

*Proof.* Induction on $\mathcal{R}$. Apart from the trivial case of $\mathcal{R}$-ID-RESOLVED, the only rule that adds a $f(w)$ to the forward configuration is $\mathcal{R}$-CHAIN-UPDATE.

Assume the property holds between $cn_F = cn'_F \, f(\text{chain } f') \, f'(w)$ and $cn_D$. We now consider $cn'_F \, f(w) \, f'(w) \mathcal{R} cn_D$.

From lemma 1 applied in $cn_F$, $f(f') \in cn_D$. And from the induction hypothesis, there exists $f'_0(f'_1)\ldots f'_{n-1}(f'_n) \in cn_D$ such that $f' \equiv f'_0$, $f'_n(w) \in cn_D$ and $\forall i, f'_i(w) \in cn_F$.

The sequence $(f_k)$ with $f_0 = f$, and $f_i = f'_{i-1}$ for $1 \leq i \leq n+1$, is such that $f_0 = f$, $f_n(w) \in cn_D$, and $\forall i, f_i(w) \in cn_F$. Thus the induction hypothesis still holds for $cn'_F \, f(w) \, f'(w) \mathcal{R} cn_D$. □

**Lemma 3** (Resolved future in DeF). *If $cn_F \mathcal{R} cn_D$ and $f(w) \in cn_D$, then:*
- *Either $w$ is a future and $f(\text{chain } w) \in cn_F$.*
- *Or there exists $w'$ such that $f(w') \in cn_F$ and there exists $f_0(f_1)\ldots f_{n-1}(f_n) \in cn_D$ such that $f \equiv f_0$, $f_n(w') \in cn_D$, and $\forall i, f_i(w') \in cn_F$.*

*Proof.* By definition of $\mathcal{R}$ and case analysis on the possible configurations, there exists $w'$ such that either $f(\text{chain } w') \in cn_F$ or $f(w') \in cn_F$. Then we conclude using lemmas 1 and 2. □

**Lemma 4** (Matching tasks). *If $cn_F \mathcal{R} cn_D$, then $\exists s. f(\{l \mid s\}\#\overline{q}) \in cn_F$ if and only if $\exists s'. f(\{l \mid s'\}\#[\![q]\!]_{fwdElim}) \in cn_D$. In this case:*
- *Either $s' = [\![s]\!]_{fwdElim}$.*
- *Or $s$ is of the form $y = \text{get} * w; s_1$, and $s'$ of the form $y = \text{get} * w'; [\![s_1]\!]_{fwdElim}$, with*

$$\exists w_0 \ldots w_n \in cn_F. \, \exists w'_0 \ldots w'_m \in cn_D \begin{cases} \forall i < n \; w_i(w_{i+1}) \in cn_F \\ \forall i < m \; w'_i(w'_{i+1}) \in cn_D \\ w_0 \equiv w'_0 \\ w_n = w \\ w'_m = w' \end{cases}$$

*Proof.* An induction on $\mathcal{R}$ proves this property. Essentially, the rule $\mathcal{R}$-Get-Future-F adds one more $w_i$, the rule $\mathcal{R}$-Get-Future-D adds one more $w'_i$, and the other cases are trivial.

We detail the case of $\mathcal{R}$-Get-Future-D. Suppose the lemma holds for
$cn_F \ f''(\{\ell \mid s\}\#\overline{q}) \ \mathcal{R} \ cn_D \ f(f') \ f''(\{\ell \mid y = \mathtt{get}* \ f; s'\}\#[\![\overline{q}]\!]_{fwdElim})$, and consider
$cn_F \ f''(\{\ell \mid s\}\#\overline{q}) \ \mathcal{R} \ cn_D \ f(f') \ f''(\{\ell \mid y = \mathtt{get}* \ f'; s'\}\#[\![\overline{q}]\!]_{fwdElim})$.

Two cases appear from the induction hypothesis.

- If $(y = \mathtt{get}* \ f \ ; \ s') = [\![s]\!]_{fwdElim}$, then $s$ is of the form $y = \mathtt{get}* \ f \ ; \ s_1$ with $s_1$ such that $s' = [\![s_1]\!]_{fwdElim}$. In this case our new pair of configurations will verify the second case of the lemma with the sequence $(w_i)$ reduced to $(f)$ and $(w'_i) = (f, f')$.
- Otherwise, $s$ is of the form $y = \mathtt{get}* \ w \ ; \ s_1$, and there exist $w_0 \ldots w_n$ and $w'_0 \ldots w'_m$ such that

$$\begin{cases} \forall i < n \ w_i(w_{i+1}) \in cn_F \ f''(\{\ell \mid s\}\#\overline{q}) \\ \forall i < m \ w'_i(w'_{i+1}) \in cn_D \ f(f') \ f''(\{\ell \mid y = \mathtt{get}* \ f; s'\}\#[\![\overline{q}]\!]_{fwdElim}) \\ w_0 \equiv w'_0 \\ w_n = w \\ w'_m = f \end{cases}$$

The sequences $w_0 \ldots w_n$ and $w'_0 \ldots w'_m$ satisfy the lemma for
$cn_F \ f''(\{\ell \mid s\}\#\overline{q}) \ \mathcal{R} \ cn_D \ f(f') \ f''(\{\ell \mid y = \mathtt{get}* \ f'; s'\}\#[\![\overline{q}]\!]_{fwdElim})$ □

**Lemma 5** (Sequence of futures: DeF to DeF+F).
*If $cn_F \mathcal{R} cn_D$ and $f_0(f_1) \ldots f_{n-1}(f_n) \ f_n(w) \in cn_D$ with $\nexists f \in cn_D. w = f$, then there exists $k_0 \leqslant \cdots \leqslant k_l$ such that for all $i < l$ either $f_{k_i}(f_{k_{i+1}}) \in cn_F$ or $f_{k_i}(\mathtt{chain} \ f_{k_{i+1}}) \in cn_F$ with $k_0 = 0$ and $k_l = n$.*

*Proof.* We build a strictly increasing sequence of indices $(k_i)$ bounded by $n$.

Suppose we have $k_0 \leqslant \cdots \leqslant k_j$ such that $\forall i < j$ either $f_{k_i}(f_{k_{i+1}}) \in cn_F$ or $f_{k_i}(\mathtt{chain} \ f_{k_{i+1}}) \in cn_F$, and $k_0 = 0$, $k_j < n$. Two cases arise from lemma 3 applied to $f_{k_j}$:

- Either $f_{k_j}(\mathtt{chain} \ f_{k_j+1}) \in cn_F$, in which case we can take $k_{j+1} = k_j + 1$.
- Or there exists $w'$ such that $f_{k_j}(w') \in cn_F$ and $f'_0(f'_1) \ldots f'_{m-1}(f'_m) \in cn_D$ such that $f'_0 = f_{k_j}$ and $f'_m(w') \in cn_D$. It is clear that $f'_i = f_{k_j+i}$. As $\nexists f.w = f$, this implies that $k_j + m \leqslant n$, thus we can take $k_{j+1} = k_j + m \geqslant k_j$.

The resulting integer sequence is strictly increasing and bounded by $n$, which proves the lemma. □

**Lemma 6** (Sequence of futures: DeF+F to DeF).
*If $cn_F \mathcal{R} cn_D$ and $f_0(f_1) \ldots f_{n-1}(f_n) \ f_n(w) \in cn_F$,
then there exists $f'_0(f'_1) \ldots f'_{l-1}(f'_l) \in cn_D$ such that $f'_0 = f_0$ and $f'_l(w) \in cn_D$.*

*Proof.* For each $i$ ($0 \leqslant i < n$), lemma 2 proves that there exists a sequence of futures from $f_i$ to $f_{i+1}$. We choose the sequence $(f')$ as the concatenation of these sequences. □

**A.2 Main Bisimulation Theorem**

**Theorem 1** (Correctness of the translation from DeF+F to DeF). *$\mathcal{R}$ is a branching bisimulation between the operational semantics of the DeF+F program P and the operational semantics of the DeF program $[\![P]\!]_{fwdElim}$.*

Let R range over observable transitions. If $cn_F \mathrel{\mathcal{R}} cn_D$ then:

$$cn_F \xrightarrow{\tau}{}^* cn'_F \implies cn'_F \mathrel{\mathcal{R}} cn_D \qquad cn_D \xrightarrow{\tau}{}^* cn'_D \implies cn_F \mathrel{\mathcal{R}} cn'_D$$
$$cn_F \xrightarrow{R} cn'_F \implies \exists cn'_D. cn_D \xrightarrow{\tau}{}^* \xrightarrow{R} cn'_D \wedge cn'_F \mathrel{\mathcal{R}} cn'_D$$
$$cn_D \xrightarrow{R} cn'_D \implies \exists cn'_F. cn_F \xrightarrow{\tau}{}^* \xrightarrow{R} cn'_F \wedge cn'_F \mathrel{\mathcal{R}} cn'_D$$

*The transitions GET-FUTURE and CHAIN-UPDATE are non-observable, both of them are labelled $\tau$. The observable transitions FORWARD-ASYNC and FORWARD-DATA are labelled RETURN-ASYNC, and FORWARD-SYNC is labelled RETURN-SYNC. All the other transitions are labelled with their original rule name.*

*Proof of theorem 3.* From the form of the theorem, we have to prove one implication per rule of the operational semantics of DeF + F, and one implication per rule of the operational semantics of DeF. We start with the only $\tau$-rule of DeF: GET-FUTURE.

**Case of $R$ = GET-FUTURE.** We suppose $cn_D \xrightarrow{\text{GET-FUTURE}} cn'_D$ and $cn_F \mathrel{\mathcal{R}} cn_D$. We have to prove $cn_F \mathrel{\mathcal{R}} cn'_D$. The definition of GET-FUTURE implies that
$f(\{\ell \mid y = \text{get} * w ; [\![s]\!]_{fwdElim}\} \# \overline{[\![q]\!]_{fwdElim}}) \in cn_D$, $w(w') \in cn_D$, and
$f(\{\ell \mid y = \text{get} * w' ; [\![s]\!]_{fwdElim}\} \# \overline{[\![q]\!]_{fwdElim}}) \in cn'_D$.
The rule $\mathcal{R}$-GET-FUTURE-D of $\mathcal{R}$ immediately gives $cn_F \mathrel{\mathcal{R}} cn'_D$.

We now suppose $cn_D \xrightarrow{R} cn'_D$ and $cn_F \mathrel{\mathcal{R}} cn_D$, with $R$ an observable rule of DeF, and we aim to prove that there exists a $cn'_F$ such that $cn_F \xrightarrow{\tau}{}^* \xrightarrow{R} cn'_F$ and $cn'_F \mathrel{\mathcal{R}} cn'_D$. We will detail the simple case of SKIP, then move on to the two interesting non-trivial cases: RETURN-ASYNC and GET-DATA.

**Case of $R$ = SKIP.** The form of the rule SKIP implies that
$cn_D = cn''_D \, f(\{\ell \mid \text{skip} ; [\![s]\!]_{fwdElim}\} \# \overline{[\![q]\!]_{fwdElim}})$ and
$cn'_D = cn''_D \, f(\{\ell \mid [\![s]\!]_{fwdElim}\} \# \overline{[\![q]\!]_{fwdElim}})$.

From lemma 4, $cn_F = cn''_F \, f(\{\ell \mid \text{skip} ; s\} \# \overline{q})$. Take $cn_F \xrightarrow{\text{SKIP}} f(\{\ell \mid s\} \# \overline{q}) \, cn''_F$, and $cn'_F \mathrel{\mathcal{R}} cn'_D$ remains to be proven.

This can be done by induction on $\mathcal{R}$. Indeed we can build a proof tree for $cn'_F \mathrel{\mathcal{R}} cn'_D$ with the same shape as the one of $cn_F \mathrel{\mathcal{R}} cn_D$, the only difference being that the application of the rule $\mathcal{R}$-FORWARDELIM that introduces $f(\{\ell \mid \text{skip} ; s\} \# \overline{q})$ is replaced by an application of the same rule but for $f(\{\ell \mid s\} \# \overline{q})$ instead.

**Case of $R$ = RETURN-ASYNC.** $a \,\rangle\, F \, f(\{\ell \mid \text{return } v\}) \xrightarrow{\text{RETURN-ASYNC}} a \,\rangle\, F \, f(w)$. If $cn_F$ is of the form $cn''_F \, f(\{\ell \mid \text{return } v\})$, then we can take $cn_F \xrightarrow{\text{RETURN-ASYNC}} cn''_F \, f(w)$. Otherwise, $cn_F$ is of the form $cn''_F \, f(\{\ell \mid \text{forward} * v\})$, in which case we can take $cn_F \xrightarrow{\text{FORWARD-ASYNC}} cn''_F \, f(\text{chain } v)$ and the rule $\mathcal{R}$-FORWARD-ASYNC gives $cn'_F \mathrel{\mathcal{R}} cn'_D$.

**Case of $R$ = GET-DATA.** We suppose $cn_D \xrightarrow{\text{GET-DATA}} cn'_D$. This implies that $cn_D$ contains a term of the form $f(\{\ell \mid y = \text{get} * w' \, ; \, [\![s]\!]_{\textit{fwdElim}}\} \# \overline{[\![q]\!]_{\textit{fwdElim}}})$, and $cn'_D$ contains a term of the form $f(\{\ell \mid y = w' \, ; \, [\![s]\!]_{\textit{fwdElim}}\} \# \overline{[\![q]\!]_{\textit{fwdElim}}})$ such that $\nexists f \in cn_D . w' = f$.

From lemma 4, $f(\{\ell \mid y = \text{get} * w \, ; \, s\} \# \overline{q}) \in cn_F$ and

$$\exists w_0 \ldots w_n \in cn_F . \, \exists w'_0 \ldots w'_m \in cn_D \begin{cases} \forall i < n \; w_i(w_{i+1}) \in cn_F \\ \forall i < m \; w'_i(w'_{i+1}) \in cn_D \\ w_0 \equiv w'_0 \\ w_n = w \\ w'_m = w' \end{cases}$$

Then, lemma 5 applied to $(w'_0, \ldots w'_m)$ gives us $k_0 \leqslant \cdots \leqslant k_l$ such that

$\forall i < l$ either $w'_{k_i}(w'_{k_{i+1}}) \in cn_F$ or $w'_{k_i}(\text{chain } w'_{k_{i+1}}) \in cn_F$

with $k_0 = 0$ and $k_l = m$.

$(w'_{k_0} \ldots w'_{k_l})$ and $(w_0 \ldots w_n)$ are both sequences of futures in $cn_F$, and they start from the same future $w_0$. Since a future can be defined only once in a configuration, these sequences coincide. As by definition the sequence $(w'_{k_0} \ldots w'_{k_l})$ cannot be further extended at its right, necessarily $n \leqslant l$. This results in: $(w'_{k_0} \ldots w'_{k_n}) = (w_0 \ldots w_n)$. Also, $w = w_n = w'_{k_n}$.

This means that $a \rangle F \; f(\{\ell \mid y = \text{get} * w \, ; \, s\} \# \overline{q}) \xrightarrow{\tau^{l-n}} a \rangle F \; f(\{\ell \mid y = \text{get} * w' \, ; \, s\} \# \overline{q})$, with the $\tau$-transitions a series of GET-FUTURE and CHAIN-UPDATE. This proves that $a \rangle F \; f(\{\ell \mid y = \text{get} * w \, ; \, s\} \# \overline{q}) \xrightarrow{\tau^{l-n}} \xrightarrow{\text{GET-DATA}} a \rangle F \; f(\{\ell \mid y = w' \, ; \, s\} \# \overline{q})$, which concludes this case.

We now consider the simulation of a transition on the DeF+F side. We first consider the $\tau$ case: suppose $cn_F \xrightarrow{\tau} cn'_F$ and $cn_F \, \mathcal{R} \, cn_D$. This side has two $\tau$-transitions.

**Case of $R$ = CHAIN-UPDATE.** $a \rangle F \; f(\text{chain } f') \; f'(w) \xrightarrow{\text{CHAIN-UPDATE}} a \rangle F \; f(w) \; f'(w)$. The rule $\mathcal{R}$-CHAIN-UPDATE allows to conclude immediately.

**Case of $R$ = GET-FUTURE.** Similar to the case of GET-FUTURE on the DeF side, but with $\mathcal{R}$-GET-FUTURE-F instead of $\mathcal{R}$-GET-FUTURE-D.

We now suppose $cn_F \xrightarrow{R} cn'_F$ and $cn_F \, \mathcal{R} \, cn_D$ with $R$ an observable rule of DeF+F, and we aim to prove that there exists a $cn'_D$ such that $cn_D \xrightarrow{\tau}{}^* \xrightarrow{R} cn'_D$ and $cn'_F \, \mathcal{R} \, cn'_D$. Once again, most cases are trivial. We thus only detail the non-trivial cases: FORWARD-ASYNC and GET-DATA.

**Case of $R$ = FORWARD-ASYNC.** We have $[\![v]\!]_{a+\ell} = f'$ and
$a \rangle F \; f(\{\ell \mid \text{forward} * v \, ; \, s'\}) \xrightarrow{\text{FORWARD-ASYNC}} a \rangle F \; f(\text{chain } f')$. From lemma 4, $f(\{\ell \mid \text{return } v \, ; \, s\}) \in cn_D$ with $(\text{forward} * v \, ; \, s') = [\![\text{return } v \, ; \, s]\!]_{\textit{fwdElim}}$. This allows to apply RETURN-ASYNC on the DeF side:

$a \; f(\{\ell \mid \text{return } v \, ; \, s\}) \rightarrow a \; f(f')$

Then the rule $\mathcal{R}$-FORWARD-ASYNC allows to conclude.

**Case of $R$ = Get-Data.** We suppose $cn_{\text{F}} \overset{\text{Get-Data}}{\to} cn'_{\text{F}}$. This implies that $cn_{\text{F}}$ contains a term of the form $f(\{\ell \mid y = \text{get*} \ w \ ; \ s\}\#\overline{q})$, and $cn'_{\text{F}}$ contains a term of the form $f(\{\ell \mid y = w \ ; \ s\}\#\overline{q})$.

From lemma 4, $f(\{\ell \mid y = \text{get*} \ w' \ ; \ [\![s]\!]_{\textit{fwdElim}}\}\#\overline{[\![q]\!]_{\textit{fwdElim}}}) \in cn_{\text{D}}$ and

$$\exists w_0 \ldots w_n \in cn_{\text{F}}. \ \exists w'_0 \ldots w'_m \in cn_{\text{D}} \begin{cases} \forall i < n \ w_i(w_{i+1}) \in cn_{\text{F}} \\ \forall i < m \ w'_i(w'_{i+1}) \in cn_{\text{D}} \\ w_0 \equiv w'_0 \\ w_n = w \\ w'_m = w' \end{cases}$$

Then, lemma 6 gives us $f''_0(f''_1)\ldots f''_{l-1}(f''_l) \in cn_{\text{D}}$ such that $f''_0 = w_0$ and $f''_l(w) \in cn_{\text{D}}$.

From their definition, it is clear that $(f''_0 \ldots f''_{m-1}) = (w'_0 \ldots w'_{m-1})$, and $m \leqslant l$. Also, $w' = w'_m = f''_m$.

This means that

$$a \rangle F \ f(\{\ell \mid y = \text{get*} \ w' \ ; \ [\![s]\!]_{\textit{fwdElim}}\}\#\overline{[\![q]\!]_{\textit{fwdElim}}}) \overset{\tau^{l-m}}{\to}$$
$$a \rangle F \ f(\{\ell \mid y = \text{get*} \ w \ ; \ [\![s]\!]_{\textit{fwdElim}}\}\#\overline{[\![q]\!]_{\textit{fwdElim}}})$$

with the $\tau$-transitions a series of Get-Future. This proves that

$$a \rangle F \ f(\{\ell \mid y = \text{get*} \ w' \ ; \ [\![s]\!]_{\textit{fwdElim}}\}\#\overline{[\![q]\!]_{\textit{fwdElim}}}) \overset{\tau^{l-m}}{\to}\overset{\text{Get-Data}}{\to}$$
$$a \rangle F \ f(\{\ell \mid y = w \ ; \ [\![s]\!]_{\textit{fwdElim}}\}\#\overline{[\![q]\!]_{\textit{fwdElim}}})$$

which concludes this case. □

## B Typing and Implementing forward*

In this subsection, we present how an implementation of forward* can be typed. As a mean of comparison, we will consider the forward construct in the Encore programming language, which is similar to ours.

The formal typing rule T-Forward given in figure 4 imposes that a function containing forward* $x$ must be typed as $t_1..t_n \to \text{Flow}[t]$.

If we look at the Encore language, a function containing forward $x$, with $x :: \text{Fut}[t]$, is typed $t_1..t_n \to t$. Calling the function asynchronously actually returns a Fut[]. This is in effect an alternate way to type forward. To add the notion of Flow[] from DeF+F to the Encore language, one should consider a forward* operator, akin to the existing forward* but operating on Flow[].

There are now two possible ways to type forward*, the DeF+F way called *strict* mode in the paper, and the *flexible* mode actually used for the implementation of forward* in the Encore compiler. We will now discuss those two typing solutions.

Nicolas Chappe, Ludovic Henrio, Amaury Maillé, Matthieu Moy, and Hadrien Renaud

### B.1 Flexible vs Strict Typing of Future

**The Strict Way**  The first possibility is what we have introduced previously in DeF+F (figure 4). Coding a method with strict typing would look like:

```
1   def work(arg: s): Flow[t]
2     forward*(worker!compute(arg))
3   end
```

Notice the presence of Flow in the type of the result of the method. If the return type was int, the user could trigger an implicit synchronisation on a synchronous call to the function, for example by this.work(arg). Returning a future forces the user to trigger explicitly synchronisations, without being burdensome for the user as only one data-flow synchronisation is needed.

Typing forward* in Encore this way doesn't require more typing or semantic rules than what we have already established.

**The Flexible Way**  This way is used in the Encore compiler to type its own forward construct. In Encore, forward works on a Fut[t]. The return type of a function using Encore's forward is then $t$, as seen in the following code:

```
1   def work(arg: s): t
2     forward worker!compute(arg)
3   end
```

Notice the absence of Fut[] in the function return type. If work is called asynchronously there no problem, but if work is called synchronously, this may hide the fact that the result of work is still an asynchronous computation. In the type system of DeF+F, this would correspond to the rule:

$$\frac{\Gamma \vdash e : \texttt{Flow}[T'] \qquad \Gamma(\texttt{m}) = \overline{T} \to T'}{\Gamma \vdash_{\texttt{m}} \texttt{forward}*\ e} \quad \text{T-CeF-Forward}$$

Then the problem is to give a semantics to the use of forward* inside a synchronous call. In principle, a function that performs a forward* is made to be invoked asynchronously but it is always possible to invoke it synchronously. Figure 6 describes the semantic rule that should replace the Forward-Sync rule of DeF+F (figure 2) in the flexible solution. Consider a synchronous invocation to a function that finishes with a forward*. As the invoker of the function cannot know that the result should be Flow[T], it is necessary to synchronise the returned value before returning it (with the typing of figure 2 the invoker is aware that the result is a Flow[T]) and ensure the returned value is not a Flow anymore.

CeF-Forward-Sync

$$\frac{[\![v]\!]_{a+\ell} = w \qquad y\ \text{fresh variable}}{a \mathrel{\rangle} F\ f(\{\ell' \mid \texttt{forward}* \ v\ ;\ s\}\#q\#\overline{q}) \to a \mathrel{\rangle} F\ f(\{\ell' \mid y = \texttt{get}* \ w\ ;\ \texttt{return}\ y\ ;\ s\}\#q\#\overline{q})}$$

■ **Figure 6** Semantic rule of a synchronous forward*.

### B.2 Benefits and Drawbacks

**Synchronous Calls**   The main argument against flexible typing is that it induces an implicit get synchronisation when a method using forward* is called in a synchronous way, as shown in the semantic rule in figure 6. This drawback only applies to methods, as forward cannot be used inside functions. Consider the following example that uses forward with the flexible typing of Encore.

```
1  active class A
2    def print_job_result(arg: s): t
3      val result = this.work(arg)
4      println("Result: {}", result)
5      return result
6    end
7    def work(arg: s): t
8      forward worker!compute(arg)
9    end
10 end
```

Here a call to print_job_result will implicitly block the actor on line val result = this.work(arg). This behaviour is similar to the way implicit futures work: method work, written with futures in mind since it uses forward*, can be called transparently like a method of a non-active class.

With the strict typing, the user will have to write an explicit get to resolve the result of the synchronous call. This is expected from a language with explicit futures.

An example of misleading behaviour induced by the flexible typing is illustrated in listing 3, while listing 4 shows how strict typing can make such behavior avoidable. In listing 3, bar uses forward hence it is naturally called asynchronously. However, Encore allows the programmer to use the synchronous call syntax this.bar() as a shortcut for get(this!bar()), which triggers a deadlock since the instance of Foo is still processing the message that triggered the invocation of foo. In order for the message bar to be processed, foo needs to finish, and foo can only finish once the bar message is processed. While the shortcut may make the code more concise and allow synchronous-like syntax, it also makes the get synchronisation implicit, which is misleading when using explicit futures. On the other hand, in listing 4, because the typing of forward* is strict, bar explicitly returns a Flow, and the synchronous call syntax induces an effectively synchronous call that returns a Flow on which the caller needs to explicitly call get to get the actual value. The potential deadlock is explicit in this version, and can be avoided by adding an await* statement that waits until the future is resolved, allowing baz to execute in the meantime. Thanks to the collapse rule, calling a method returning a Flow asynchronously doesn't create any nested flow (unlike what would happen with a method returning Fut), hence returning a Flow explicitly isn't a problem.

■ **Listing 3** Forward with futures in classic Encore

```
1  active class Foo
2      def foo(): int
3          this.bar() -- Types in classic Encore, but
               ↪ deadlocks
4      end
5
6      def bar(): int
7          forward(this!baz())
8      end
9
10     def baz(): int
11         42
12     end
13 end
14
15 active class Main
16     def main(): unit
17         var f = new Foo()
18         print(get*(f!foo()))
19     end
20 end
```

■ **Listing 4** Forward with flows in extended Encore

```
1  active class Foo
2      def foo(): int
3          var x = this.bar()
4          -- get*(x) -- Deadlock as there is
               ↪ already a message being
               ↪ processed
5          await*(x) -- Wait until forward* is
               ↪ completed
6          get*(x) -- Now it doesn't deadlock
7      end
8
9      def bar(): Flow[int]
10         forward*(this!!baz())
11     end
12
13     def baz(): int
14         42
15     end
16 end
17
18 active class Main
19     def main(): unit
20         var f = new Foo()
21         print(get*(f!foo()))
22     end
23 end
```

**Backward Compatibility**  The main advantage for using flexible typing is that code can seamlessly be translated from control-flow explicit futures Fut[] to data-flow explicit futures Flow[]. Let us consider the following example with control-flow explicit futures:

```
1  active class LinkedListNode
2      val state: int
3      val next: Maybe[Node]
4      -- …
5      def sum(acc: int = 0): int
6          match this.next with
7              case Just(next) => forward(next!sum(this.state + acc))
8              case _ => this.state + acc
9          end
10     end
11 end
```

To port this code to DeF+F, one can just change the asynchronous operators: forward into forward* and ! (asynchronous call returning a Fut[]) into !! (asynchronous call returning a Flow[]). This operation is very easy and does not require any major change in the code. In particular, the type system of Encore with DeF includes the previously control-flow explicit futures type system.

On the other hand, if we use strict typing, porting this code to Encore with DeF can be burdensome, as the return types of functions may change. This would require programmers to manually change the signature of all functions using forward and then track calls to these functions to change the code.

### B.3 Decision

We chose to implement the flexible version into our Encore extension, to provide backward compatibility between the DeF and the original Encore compiler. That being said, it is recommended to type the methods using forward* in a strict way, as to emphasise their asynchronous nature, and prevent synchronous calls to asynchronous methods. In the context of a fresh language developed with data-flow explicit futures, we would advise to use the strict rule.

The bisimulation proven in our paper still holds in this setting as long as forward* is only used in asynchronously called functions. The additional flexibility introduced with our implementation does not allow us to have the equivalence between forward* $v$ and return $v$ (translation $[\![]\!]_{fwdElim}$) in synchronous calls. Indeed the additional synchronisation introduced by the semantics of forward might block a program that would otherwise terminate, for example. Note that potentially problematic cases (synchronous calls to asynchronous methods) can easily be forbidden statically, hence this is not a limitation of our choice.

**About the authors**

**Nicolas Chappe** Nicolas Chappe is an internship student from Ecole Normale Supérieure de Lyon. email: nicolas.chappe@ens-lyon.fr.

**Ludovic Henrio** Ludovic Henrio is a researcher in the Cash team at LIP laboratory in Lyon. email: ludovic.henrio@cnrs.fr.

**Amaury Maillé** Amaury Maillé is a PhD student in the Cash team at LIP laboratory in Lyon. email: amaury.maille@ens-lyon.fr.

**Matthieu Moy** Matthieu Moy is assistant professor in the Cash team at LIP laboratory in Lyon. email: matthieu.moy@univ-lyon1.fr.

**Hadrien Renaud** Hadrien Renaud is an internship student from Ecole Polytechnique. email: hadrien.renaud@polytechnique.edu.